\documentclass[fleqn,usenatbib]{mnras}

\usepackage{newtxtext,newtxmath,bm}

\usepackage[T1]{fontenc}

\DeclareRobustCommand{\VAN}[3]{#2}
\let\VANthebibliography\thebibliography
\def\thebibliography{\DeclareRobustCommand{\VAN}[3]{##3}\VANthebibliography}

\usepackage{graphicx}    
\usepackage{amsmath}    

\title[Improving divergence cleaning in cosmological SPMHD]{Improving divergence cleaning in cosmological SPMHD simulations}

\author[UPS and DJP]{
Ulrich P. Steinwandel$^{1}$\thanks{E-mail: uli@mpa-garching.mpg.de (UPS)}
and Daniel J. Price$^{2,3}$
\\
$^{1}$Max Planck Institute f\"ur Astrophysik, Karl-Schwarzschild Str.1, Garching, D-85748, Germany\\
$^{2}$School of Physics and Astronomy, Monash University, Clayton, VIC, 3800, Australia\\
$^{3}$IPAG, Univ. Grenoble Alpes, CNRS, 38000 Grenoble, France
}
\date{Accepted XXX. Received YYY; in original form ZZZ}

\pubyear{2026}

\begin{document}
\label{firstpage}
\pagerange{\pageref{firstpage}--\pageref{lastpage}}
\maketitle

\begin{abstract}
We implement the constrained hyperbolic/parabolic divergence cleaning algorithm into the cosmological smoothed particle magnetohydrodynamics (SPMHD) code {\sc OpenGadget3}, modifying the governing equations so that the scheme can be applied consistently in an expanding cosmological framework. This ensures that divergence errors in the magnetic field are actively propagated away and damped, rather than advected with the flow and partially controlled by source terms as in the previously employed Powell eight-wave approach. We validate the implementation on a series of standard test problems --- the advection of an artificial divergence pulse, the Orszag--Tang vortex, the Brio--Wu shock tube, and a magnetised Zeldovich pancake --- which confirm that the scheme reduces divergence errors while preserving the correct physical evolution. We then apply the method to a fully cosmological simulation of a massive galaxy cluster with $M_\mathrm{200c} \approx 10^{15}~\mathrm{M}_{\odot}$, and compare directly to the Powell-only approach. The overall density structure of the cluster is largely unchanged by the choice of divergence cleaning, and the magnetic field geometry and strength in the cluster core remain similar. In the cluster outskirts ($r \approx 1$--$3~h^{-1}$~Mpc), however, the magnetic field is amplified by a factor of 5--10 compared to the Powell-only run, while the divergence error is reduced by 2--3 orders of magnitude throughout the cluster volume.
Our results suggest that accurate divergence control is essential for capturing magnetic field amplification in the low-density, poorly resolved outskirts of galaxy clusters.
\end{abstract}




\section{Introduction}
\label{sec:intro}
Our Universe is magnetic. Magnetism extends from the largest structures: galaxies, and the intra-cluster medium (ICM) of galaxy clusters, down to the smallest scales: the interstellar medium (ISM), stars and planets. While magnetic field strengths in stars range from several Gauss (G) in the Sun to $10^{15}$ G in pulsars, on larger scales B-field strengths typically saturate at the canonical value of a few to a few tens of $\mu$G in galaxies and galaxy clusters but can reach strengths of mG in the dense ISM \citep{Crutcher2012,Pattle2023}.

Past and current research has developed the following picture of magnetic field amplification in the Universe: A small scale-turbulent dynamo amplifies tiny seed fields to the values we observe today in galaxies and galaxy clusters at $\mu$G-level. The origin of these seed fields is still under debate and several processes have been suggested that can generate seed fields of the order of around $10^{-20}$ G \citep[e.g.][]{Biermann1950, Harrison1970, Demozzi2010, Gnedin2000, Durier2012}. In galaxies, these fields are ordered and influenced by a large-scale (mean-field) $\alpha$-$\Omega$ dynamo \citep[e.g][]{Parker1955, Steenbeck1966, Parker1979, Ruzmaikin1988} and can be ejected in galactic outflows that can, in turn, magnetize the circumgalactic medium (CGM) \citep[e.g.][]{Bertone2006, Pakmor2017, Pakmor2020, vandeVoort2021}.

On galaxy cluster scales the question of how the observed $\sim \mu$G fields are seeded remains open. The intra-cluster gas is strongly metal-enriched by outflows from member galaxies, so it is reasonable to expect it to be magnetically enriched by the same outflows as well \citep[e.g.][]{Bertone2006, Pakmor2017, Pakmor2020, vandeVoort2021}. In contrast, primordial-only seeds at the canonical Biermann-battery level ($\sim 10^{-20}$~G) are not amplified to $\mu$G by structure-formation turbulence and merger shocks alone in current cosmological simulations. To isolate the ICM dynamo from the computationally expensive physics of galaxy-scale outflows, previous numerical studies of cluster magnetic-field amplification have therefore adopted elevated initial seed fields, chosen so that the field saturates at $\sim \mu$G in the cluster centre by redshift zero \citep[e.g.][]{Vazza2018, Steinwandel2022_cluster}. Reported seed-field values span several orders of magnitude --- e.g.\ $\sim 10^{-14}$~G comoving at $z \approx 100$ in \citet{Steinwandel2022_cluster}, $\sim 10^{-8}$~G in \citet{Vazza2018} --- but the strategy is the same: an enhanced, idealised seed substitutes for the missing wind-injection channel and allows the dynamo to be studied in isolation. We adopt this same testing strategy in the present work, since our focus is the numerical performance of the divergence-cleaning scheme rather than the seed-field problem. Even with this idealisation, the resulting fields will be ordered on larger scales by the structure formation process, with some evidence that the void magnetic field can ``remember'' the structure of the initial seed field on the scales of a few 10~Mpc \citep[e.g.][]{Mtchedlidze2021}.

The central process behind the turbulent amplification of magnetic fields is the stretch-twist-fold mechanism, introduced by \citet{Zeldovich1970} and researched by a number of groups \citep[e.g.][]{Kraichnan1967, Kazantsev1968, Kazantsev1985, Kulsrud1992, Brandenburg1995, Kulsrud1997, Xu2020}.
On both galaxy and galaxy cluster scales simulations using a variety of methods have established that this process is dominant in amplifying magnetic fields  \citep[e.g.][]{Dolag1999,Dolag2002,Kotarba2009,Dubois2008, Wang2009, Pakmor2013, Pakmor2017, Butsky2017, Garaldi2020, Steinwandel2019, Steinwandel2020, Steinwandel2022, Steinwandel2022_cluster, Vazza2014, Vazza2018}. However, small-scale dynamos can only generate correlated fields on the scale of the turbulence and require a process to order the field on the largest scales. For instance, one can derive the outer scale of MHD turbulence from the peak in the magnetic energy spectra, set by the magnetic Reynolds number which compares the advection timescale to the magnetic diffusion timescale.

Of concern in current MHD simulations of the small-scale dynamo in galaxies and galaxy clusters is the degree to which the $\nabla\cdot\mathbf{B} = 0$ condition from Maxwell's equations is enforced. Standard schemes such as constrained transport \citep{EvansHawley1988} cannot easily be ported to the Lagrangian particle codes used for cosmological structure formation, including the {\sc Gadget} code family (\citealt{Springel2005}; see also {\sc gizmo}). Since the magnetic field in cosmological simulations is dynamically weak, errors in $\nabla\cdot\mathbf{B}$ are unlikely to dramatically alter the bulk matter distribution. A common practical response has been to rely on the \citet{Powell1999} 8-wave scheme, which advects divergence errors with the fluid bulk velocity but does not actively remove them. In the moving-mesh code {\sc arepo} \citep{Springel2010} the Powell scheme has been adopted in cosmological MHD simulations and appears to remain numerically stable in that setting \citep[e.g.][]{Pakmor2011, Pakmor2013}, plausibly because the higher-order gradient reconstruction available on the moving Voronoi mesh keeps the residual $\nabla\cdot\mathbf{B}$ smaller than in standard SPH. In SPH-based cosmological codes, however, residual $\nabla\cdot\mathbf{B}$ errors can reach order unity in the $B$-field \citep[e.g.][]{Kotarba2009, Beck2016, Steinwandel2019, Steinwandel2020, Steinwandel2022_cluster}, with the associated risk of unphysical $B$-field amplification \citep{Kotarba2009}.

The hyperbolic/parabolic divergence cleaning scheme proposed by \citet{Dedner2002} is straightforward to adapt to smoothed particle magnetohydrodynamics (SPMHD; Dedner cleaning was first investigated by \citealt{Price2005}). However, the stability of the original Dedner formulation in dynamic, individual-timestep settings has been a recurring difficulty in SPMHD implementations \citep{Price2012}, and is especially pronounced in the {\sc Gadget} family in the context of galactic dynamo simulations \citep{Steinwandel2019, Steinwandel2020}. \citet{Stasyszyn2013} implemented a Dedner-type cleaning in {\sc Gadget} and reported improved results in galaxy cluster simulations, but their specific choice of numerical operators (which is crucial; see \citealt{Tricco2012}) was not documented, and subsequent attempts to use the vector potential \citep{Stasyszyn2015} suggest that the scheme remained unsatisfactory.

The stability problem was resolved by \citet{Tricco2012}, who reformulated the cleaning equations so as to guarantee energy conservation in the absence of dissipation. \citet{Tricco2016} subsequently extended this stability guarantee to variable cleaning speeds, removing the need for the restrictive global timestep constraint of the original scheme: one can set the cleaning speed to the local maximum signal speed for each particle or grid cell with no additional constraint on the timestep. While their scheme was formulated with SPMHD in mind, the principles are general and have been successfully applied in grid-based codes as well \citep{Derigs2018,Mueller2020,Varma2021}. 

Constrained hyperbolic/parabolic cleaning has subsequently been applied extensively in SPMHD studies of magnetised star and planet formation, enabling stable calculations of magnetised jets \citep{PriceTriccoBate2012, BateTriccoPrice2014, Wurster2018} that would previously explode \citep{Tricco2012}, of non-ideal MHD \citep[e.g.][]{Wurster2014, Wurster2016, Wurster2017, Wurster2018, Wurster2019}, of magnetic fields in isolated galaxies \citep{Dobbs2016}, and --- most relevant to this paper --- in a code comparison of the small-scale dynamo \citep{TriccoPriceFederrath2016}. While the algorithm is the default cleaning scheme in the {\sc phantom} SPMHD code \citep{Price2018}, it has not previously been fully implemented in the {\sc Gadget} family of cosmological SPMHD codes, and has therefore not been applied at the largest scales of cosmic structure formation. The aim of this paper is to close that gap.

The main result presented here is a simulation of a massive galaxy cluster with a total mass of $2 \times 10^{15}$~M$_{\odot}$ modelled with and without active cleaning of divergence errors in the magnetic field. We study the amplification of the field and assess the effect of magnetic fields on the thermal pressure profile of the cluster. Our main aim is to test the importance of active control of divergence errors in such simulations.

The paper is structured as follows: In Sec.~\ref{sec:methods}, we discuss our implementation of the constrained cleaning scheme in {\sc OpenGadget3}. Appendix~\ref{sec:tests} gives the results of test problems to ensure our scheme is working. Our main results (Sec.~\ref{sec:results}) are cosmological simulations of a massive galaxy cluster with and without the newly implemented cleaning scheme. We discuss in Sec.~\ref{sec:discussion} and conclude in Sec.~\ref{sec:conclusions}.


\section{Methods}
\label{sec:methods}

\subsection{Simulation Code}

The end goal of this paper is to implement the constrained, energy-conserving Dedner-type cleaning scheme of \citet{Tricco2012} and \citet{Tricco2016} into the cosmological simulation code {\sc OpenGadget3} \citep{Springel2005, Beck2016, Groth2023}. The code is the developer base of P-Gadget-2/3 \citep{Springel2005} and is a Tree-multi-method code that treats hydrodynamics via SPH in density-entropy and pressure-entropy formulations as well as via the meshless finite mass (MFM) method \citep{Groth2023}. Time-dependent shock dissipation follows the SPH improvements of \citet{Beck2016}, in which the dissipation switches respond to local flow indicators rather than being held fixed. The code also solves the MHD equations in SPMHD using a density-entropy formulation. Specifically, the code solves the equations of ideal MHD:
\begin{align}
\frac{{\rm d}\rho}{{\rm d} t} & = -\rho(\nabla\cdot\mathbf{v}), \label{eq:conti} \\
\frac{{\rm d}\mathbf{v}}{{\rm d} t} & = -\frac{\nabla P}{\rho} + \mathbf{g} + \frac{(\mathbf{J} \times \mathbf{B})}{\rho}, \\
\frac{{\rm d}\mathbf{B}}{{\rm d}t} & = (\mathbf{B} \cdot \nabla) \mathbf{v} - \mathbf{B} (\nabla \cdot \mathbf{v}), \label{eq:induction} \\
\frac{{\rm d}K}{{\rm d} t} & = \frac{\gamma-1}{\rho^{\gamma-1}} \Lambda_{\rm diss}, \label{eq:entrop} \\
\nabla \cdot \mathbf{B} & = 0,
\end{align}
where ${\rm d}/{\rm d}t \equiv \partial/\partial t + \mathbf{v}\cdot\nabla$ is the Lagrangian time derivative, $\rho$ is the density, $\mathbf{v}$ is the velocity, $P$ is the pressure, $\mathbf{J}$ is the current, $\mathbf{B}$ is the magnetic field, and $K \equiv P/\rho^\gamma$ is the entropy variable. The term $\Lambda_{\rm diss}$ collects all numerical dissipation channels --- shock viscosity, resolution-scale conductivity at contact discontinuities, and resolution-scale resistivity at current sheets --- which feed dissipated kinetic and magnetic energy back into thermal energy. All three switches are time-dependent and follow the SPH implementation of \citet{Beck2016}: the local switch values rise in regions where they are needed (compressive flow, large temperature gradients, strong $\mathbf{J}$) and decay back to a small floor elsewhere. We use this scheme in its default {\sc OpenGadget3} configuration throughout this paper.

Without going into the full SPMHD framework, we draw attention to the importance of the chosen form of the induction equation. Written in Eulerian form,
\begin{align}
\frac{\partial \mathbf{B}}{\partial t} & = \nabla \times (\mathbf{v} \times \mathbf{B}) \label{eq:eulerian_mhd} \\
& = - (\mathbf{v} \cdot \nabla) \mathbf{B} + (\mathbf{B} \cdot \nabla) \mathbf{v}  - \mathbf{B} (\nabla \cdot \mathbf{v}) + \mathbf{v} (\nabla \cdot \mathbf{B}),
\end{align}
the equation is exactly equivalent to Eq.~\ref{eq:induction} in the continuum, since the final term $\mathbf{v}(\nabla\cdot\mathbf{B})$ vanishes by virtue of $\nabla\cdot\mathbf{B}=0$. The two forms differ only at the discrete level in SPMHD, because the divergence carried by the particle distribution is not identically zero. Equation~\ref{eq:induction} is the form actually solved by {\sc OpenGadget3}, and the next section shows that it is the form whose discretisation advects monopole errors with the flow.

\subsection{Divergence advection and the Powell scheme in {\sc Gadget}}

As noted above, Eqs.~\ref{eq:induction} and \ref{eq:eulerian_mhd} are equivalent in the continuum: the difference is the term $\mathbf{v}(\nabla\cdot\mathbf{B})$, which vanishes exactly when $\nabla\cdot\mathbf{B} = 0$ holds. In SPMHD the divergence carried by the particle distribution is generally non-zero, and the standard discrete operators do not satisfy the constrained-transport identity $\nabla_{\rm SPH}\cdot(\nabla_{\rm SPH}\times\,\cdot\,) \equiv 0$ that is built in to staggered-grid schemes \citep{EvansHawley1988}. The choice of induction-equation form therefore matters at the discrete level. Taking the divergence of Eq.~\ref{eq:induction} yields:
\begin{align}
    \frac{\partial}{\partial t} (\nabla \cdot \mathbf{B}) + \nabla \cdot (\mathbf{v}\,\nabla \cdot \mathbf{B}) = 0,
\end{align}
i.e.\ an advection equation for $\nabla \cdot \mathbf{B}$ with the bulk flow velocity, with the same conservative form as the continuity equation (Eq.~\ref{eq:conti}; in Eulerian form reading $\partial \rho/\partial t + \nabla\cdot(\mathbf{v}\rho) = 0$). By analogy with conservation of mass ($\int \rho {\rm d}V$), it implies that
\begin{equation}
\int (\nabla\cdot\mathbf{B}) {\rm d} V = \oint_S \mathbf{B} \cdot d\mathbf{S} = {\rm const},
\end{equation}
and hence that the global magnetic surface flux is conserved, even in the presence of $\nabla \cdot \mathbf{B}$ errors in the simulation domain. This form of the induction equation therefore advects monopole errors with the flow but does not destroy them: it is a divergence-\emph{advecting} scheme, not a divergence-cleaning one. Furthermore, every SPMHD implementation needs to address the issue of an (unphysical) source term in the momentum equation that scales as $\nabla \cdot \mathbf{B}$ \citep[see discussion in][]{Price2012}. This term typically arises because the force related to MHD is written as a divergence of the MHD stress tensor. It gives rise to the tensile instability, where the spurious source acts as an attractive force that dominates when the magnetic pressure exceeds the gas pressure \citep{Phillips1985,Morris1996,Price2012}. In {\sc Gadget}, as in other SPMHD codes (including {\sc phantom}; \citealt{Price2018}), this source term is subtracted from the momentum equation following \citet{Borve2001}, which yields an overall scheme equivalent to the \citet{Powell1999} ``8 wave scheme'' in SPH formulation. Every stable numerical SPMHD scheme therefore enforces a divergence-advecting scheme similar to the \citet{Powell1999} approach by construction.

\subsection{Constrained Dedner-type cleaning in {\sc Gadget}}
\label{sec:cleaning}

To improve on the divergence preservation built into {\sc Gadget}'s default SPMHD scheme, we implement the Dedner-type cleaning approach \citep{Dedner2002} in the constrained, energy-conserving formulation of \citet{Tricco2012}, extended to a variable cleaning speed by \citet{Tricco2016}. The classical \citet{Dedner2002} formulation has been shown to be unstable in SPMHD \citep{Tricco2012} and also in Eulerian codes \citep{BalsaraKim2004}; we therefore reproduce the key steps of the constrained derivation below, since the energy-conservation argument is directly load-bearing for our cosmological application. The reader is referred to \citet{Tricco2012} and \citet{Tricco2016} for the full development.

The starting point is the \citet{Dedner2002} idea of correcting the induction equation by the gradient of an auxiliary scalar field $\psi$ that obeys a forced, damped wave equation,
\begin{align}
    \frac{{\rm d}\mathbf{B}}{{\rm d}t} & = (\mathbf{B} \cdot \nabla) \mathbf{v} - \mathbf{B} (\nabla \cdot \mathbf{v}) - \nabla \psi, \label{eq:dedner_B}\\
    \frac{\partial \psi}{\partial t} & = -c_{h}^2 (\nabla \cdot \mathbf{B}) - \frac{\psi}{\tau}, \label{eq:dedner_psi}
\end{align}
which, after combination, yields the damped wave equation for the divergence error
\begin{align}
    \frac{\partial^2 (\nabla \cdot \mathbf{B})}{\partial t^2} - c_h^2 \nabla^2 (\nabla \cdot \mathbf{B}) + \frac{1}{\tau} \frac{\partial (\nabla \cdot \mathbf{B})}{\partial t} = 0,
    \label{eq:hyperbolic}
\end{align}
with damping time
\begin{align}
    \tau = \frac{h}{\sigma c_\mathrm{h}}, \label{eq:tau}
\end{align}
where $\sigma$ is a dimensionless numerical constant (typically $\sigma = 1$ in three dimensions; \citealt{Tricco2012}), $h$ is the local smoothing length and $c_h$ is the cleaning speed. Equation~\ref{eq:hyperbolic} shows that the divergence error is propagated as a wave at speed $c_h$ and damped on timescale $\tau$, redistributing it over a non-local region of the domain.

In its original (Eulerian) form, however, this system is not energy-conserving in SPMHD. The $-\nabla\psi$ source in Eq.~\ref{eq:dedner_B} and the $-c_h^2(\nabla\cdot\mathbf{B})$ source in Eq.~\ref{eq:dedner_psi} exchange energy between $\mathbf{B}$ and $\psi$, but there is no guarantee that this energy is conserved once discretised. \citet{Tricco2012} showed that, as a consequence, the classical \citet{Dedner2002} formulation is unstable in SPMHD at strong density jumps and free surfaces: in those regions the cleaning can act as an unphysical \emph{source} of magnetic energy --- equivalently, a source of negative entropy --- and produce exponential, unbounded growth.

\citet{Tricco2012} resolved this by re-deriving the cleaning equations directly from a global energy budget. Defining the energy stored in the cleaning field as
\begin{align}
E_\psi = \int \frac{1}{2}\,\frac{\psi^2}{\mu_0\,\rho\,c_h^2}\,\rho\,{\rm d}V,
\label{eq:Epsi}
\end{align}
the total energy $E = E_{\rm kin} + E_{\rm th} + E_{\rm mag} + E_\psi$ obeys
\begin{align}
\frac{{\rm d}E}{{\rm d}t} = -\int \frac{\psi^2}{\mu_0\,\rho\,c_h^2\,\tau}\,{\rm d}V \le 0,
\label{eq:Ebudget}
\end{align}
provided the cleaning equations are written as
\begin{align}
    \frac{{\rm d}\mathbf{B}}{{\rm d}t} & = (\mathbf{B} \cdot \nabla) \mathbf{v} - \mathbf{B} (\nabla \cdot \mathbf{v}) - \nabla \psi, \label{eq:T2012_B}\\
    \frac{{\rm d}\psi}{{\rm d}t} & = -c_\mathrm{h}^2 (\nabla \cdot \mathbf{B}) - \frac{\psi}{\tau} - \frac{1}{2}\,\psi\,(\nabla \cdot \mathbf{v}). \label{eq:T2012_psi}
\end{align}
The additional $-\tfrac{1}{2}\psi(\nabla\cdot\mathbf{v})$ source term in Eq.~\ref{eq:T2012_psi} arises naturally from the Lagrangian time derivative of $E_\psi$: the explicit $\rho^{-1}$ in the energy density combined with the continuity equation $\dot{\rho} = -\rho(\nabla\cdot\mathbf{v})$ to produce exactly this contribution, so that it cancels in the overall energy budget. With this term in place alongside the Lagrangian derivative on the left hand side of Eq.~\ref{eq:T2012_psi} the parabolic damping in Eq.~\ref{eq:Ebudget} is manifestly negative-semi-definite: the cleaning can only ever remove magnetic energy, never add it, and the instability identified by \citet{Tricco2012} is removed. Note that for vanishing $\psi$ the system reduces to the SPMHD scheme already in use in {\sc Gadget}, which is equivalent to the \citet{Powell1999} eight-wave scheme \citep{Borve2001}.

Equations~\ref{eq:T2012_B}--\ref{eq:T2012_psi} are valid only for spatially and temporally constant cleaning speed. In a cosmological SPH simulation with individual timesteps each particle should carry its own local signal speed, so $c_h$ varies in space and time. \citet{Tricco2016} showed that a naive variable $c_h$ re-introduces an unbalanced source term $\propto {\rm d}c_h/{\rm d}t$ in the energy budget which, left uncorrected, would again admit energy growth. The remedy is to evolve $\psi/c_h$ rather than $\psi$: because the cleaning energy density (Eq.~\ref{eq:Epsi}) is proportional to $(\psi/c_h)^2$, this choice of variable has no explicit $c_h$ dependence and absorbs the $\dot{c}_h$ source automatically. The bound ${\rm d}E/{\rm d}t \le 0$ is then preserved for arbitrary $c_h(\mathbf{x},t)$, with no need for an explicit prescription of how $c_h$ evolves. The resulting continuum equations, written in the comoving units used by {\sc Gadget} for cosmological integration (the comoving conversion is described below), read:
\begin{align}
    \frac{1}{a^2}\frac{{\rm d}\mathbf{B}}{{\rm d}t} & = (\mathbf{B} \cdot \nabla) \mathbf{v} - \mathbf{B} (\nabla \cdot \mathbf{v}) - \nabla \psi, \\
    \frac{1}{a^2}\frac{{\rm d}}{{\rm d}t}\left(\frac{\psi}{c_\mathrm{h}}\right) & = -c_\mathrm{h} (\nabla \cdot \mathbf{B}) - \frac{1}{\tau} \left(\frac{\psi}{c_\mathrm{h}}\right) - \frac{1}{2} \left(\frac{\psi}{c_\mathrm{h}}\right) (\nabla \cdot \mathbf{v}),
    \label{eq:evolution_grad_psi}
\end{align}
which can be discretized in SPH in comoving units for cosmological integration according to
\begin{align}
    \frac{1}{a^2}\left(\frac{{\rm d}\mathbf{B}_{i}}{{\rm d}t}\right)_{\psi} = & -\rho_{i} \sum_{j} m_j \left[\frac{\psi_i}{\Omega_i \rho_i^2} \nabla_i W_{ij}(h_i) + \frac{\psi_j}{\Omega_j \rho_j^2} \nabla_i W_{ij}(h_j) \right], \\
    \frac{1}{a^2}\frac{{\rm d}}{{\rm d}t}\left(\frac{\psi}{c_\mathrm{h}}\right)_i = & \frac{c_{\mathrm{h},i}}{\Omega_i \rho_i} \sum_j m_j (\mathbf{B}_i - \mathbf{B}_j) \cdot \nabla_i W_{ij}(h_i) - \frac{1}{\tau} \left(\frac{\psi}{c_\mathrm{h}}\right)_i \\
    & + \frac{1}{2} \left(\frac{\psi}{c_\mathrm{h}}\right)_i \sum_j m_j (\mathbf{v}_i - \mathbf{v}_j) \cdot \nabla_i W_{ij}(h_i), \label{eq:advect}
\end{align}
where $W$ is the kernel, $i$ and $j$ the indices of the particles over which we carry out the SPH sum and $h$ is the smoothing length.

The factor $1/a^2$ on the left-hand side of Eqs.~\ref{eq:evolution_grad_psi}--\ref{eq:advect} is not a physical correction to the cleaning equations themselves, but a bookkeeping convention inherited from {\sc Gadget}'s cosmological time integration. Following \citet{Springel2005} and the cosmological MHD conventions of \citet{Dolag1999} and \citet{Dolag2009_subfind}, the magnetic-field kicks are integrated using the supercomoving time variable
\begin{align}
\mathrm{d}t_{\rm mag} = \int \frac{\mathrm{d}t}{a^2},
\end{align}
which absorbs the leading cosmological scaling of the induction equation into the time variable rather than into explicit Hubble or expansion-rate terms. The code stores the time derivatives $\mathrm{d}\mathbf{B}/\mathrm{d}t$ and $\mathrm{d}(\psi/c_h)/\mathrm{d}t$ pre-multiplied by $a^2$, so that the kick step $\mathbf{B} \to \mathbf{B} + (\mathrm{d}\mathbf{B}/\mathrm{d}t)\,\mathrm{d}t_{\rm mag}$ recovers the physical evolution exactly: the explicit $a^2$ on the LHS and the implicit $1/a^2$ from $\mathrm{d}t_{\rm mag}$ cancel. The form written here therefore describes what the code actually integrates; it is equivalent to the physical-form equations of \citet{Tricco2016} once the cosmological time substitution is undone, and reduces to them in the non-cosmological limit $a \equiv 1$. We adopt
\begin{align}
c_\mathrm{h} = \sqrt{{\rm v}_A^2 +c_s^2},
\label{eq:clean_speed}
\end{align}
with the Alfv\'en speed $v_A$ and the sound speed $c_s$, i.e.\ the local fast magnetosonic speed. This is the same speed at which true MHD waves propagate, so cleaning the divergence error at $c_h$ adds no information that is not already required for the hydrodynamic Courant condition.

The hyperbolic damped-wave equation for $\nabla\cdot\mathbf{B}$ (Eq.~\ref{eq:hyperbolic}) propagates the divergence error at speed $c_h$, which sets the natural CFL constraint $\Delta t < C_{\rm courant}\,h/c_h$ for the cleaning step. In a pair-based SPMHD discretisation this constraint is automatically satisfied by the standard Courant timestep based on the SPH signal velocity $v_{\rm sig}$ \citep{Springel2005, Beck2016}, defined for an interacting pair $(i,j)$ as the sum of the local fast magnetosonic speeds along $\hat{\mathbf{r}}_{ij}$ plus a relative-velocity contribution that is active in compressive flow; we use the implementation of $v_{\rm sig}$ built into {\sc OpenGadget3}. Since by Eq.~\ref{eq:clean_speed} $c_h$ never exceeds $v_{\rm sig}$, the cleaning is intrinsically Courant-stable.

To allow the cleaning to be optionally pushed beyond the natural signal speed for testing purposes, we introduce a separate overcleaning factor $\zeta \ge 1$ that scales the cleaning speed used in the timestep criterion only, and not in the cleaning equations themselves. The combined timestep constraint is then
\begin{align}
    \Delta t = \mathrm{min}\left(\frac{C_\mathrm{courant}\,h}{v_\mathrm{sig}}, \;\frac{C_\mathrm{courant}\,h}{\zeta\,c_h}\right),
\end{align}
with $C_{\rm courant} = 0.3$ \citep{Monaghan1985}. We emphasise that the parameters $\zeta$ and $\sigma$ play distinct roles: $\sigma$ controls the parabolic damping rate via $\tau = h/(\sigma c_h)$ in Eq.~\ref{eq:tau} and is a property of the cleaning equations themselves, while $\zeta$ is purely a numerical knob applied to the cleaning timestep. With the default value $\zeta = 1$ the timestep is unchanged from the standard Courant condition, since $c_h \le v_{\rm sig}$. The runs labeled \texttt{NIFTY-clean2} use $\zeta = 2$, i.e.\ the cleaning is allowed to run at twice the local fast magnetosonic speed; this is what we mean below by ``faster cleaning than the fastest wave speed''.

\subsection{Simulations and Initial Conditions}

In this work we aim to present the initial implementation and testing of the constrained cleaning scheme of \citet{Tricco2012} and \citet{Tricco2016}, adapted for the use in cosmological simulations. The results of our implementation tests are given in Appendix~\ref{sec:tests}. We focus on the application of the scheme to massive galaxy clusters, using the NIFTY cluster as our test-bed for quantifying the scheme in a cosmological context. The initial conditions (ICs) for the cluster are taken from the MUSIC-2 sample \citep[e.g.][]{Prada2012, Sembolini2013, Sembolini2014, Biffi2014} and were subsequently used for a cluster comparison project in the NIFTY collaboration (hence the NIFTY cluster; \citealt{Sembolini2016}). The cluster has the following physical properties: The clusters mass M$_\mathrm{200c} = 10^{15}$ M$_{\odot}$ and is selected as zoom-in initial condition from a large Gpc-volume. The dark matter mass resolution is m$_\mathrm{dm} = 9.01 \times 10^{8} $ h$^{-1}$ M$_{\odot}$ and a gas resolution of m$_\mathrm{gas} = 1.9 \times 10^{8}$ h$^{-1}$ M$_{\odot}$. This makes the resolution similar to early simulations of galaxy cluster formation with the focus on magnetic field amplification \citep[e.g.][]{Dolag1999, Dolag2001, Dolag2002, Dolag2004, Dolag2005, Dolag2009, Bonafede2011, Vazza2011, Stasyszyn2010, Stasyszyn2013} or the lowest resolution runs in \citet{Steinwandel2022_cluster}, \citet{Vazza2014} and \citet{Vazza2018}. This mass resolution is ideal for testing and development as we can complete on run down to redshift zero in approximately a few hours. Ten or a hundred times higher resolution are drastically more expensive (weeks to months) \citep[see e.g.,][for simulations with much higher resolution]{Steinwandel2022_cluster, Steinwandel250X}. The background cosmology of the box follows a WMAP7 cosmology \citep[][]{Komatsu2011} with $\Omega_\mathrm{m} = 0.27$, $\Omega_\mathrm{b} = 0.0469$, $\Omega_\mathrm{\Lambda} = 0.73$, $\sigma_{8} = 0.82$, $n=0.95$ and $h=0.7$. We use \textsc{subfind} to identify haloes \citep[][]{Springel2001, Dolag2009_subfind}. For these runs we additionally include physical thermal conduction (i.e.\ the Spitzer--H\"arm electron heat flux), implemented as an isotropic conductivity of the form $\kappa = f_{\rm Sp}\,\kappa_{\rm Sp}(T)$ with $f_{\rm Sp} = 0.05$ and $\kappa_{\rm Sp}(T) \propto T^{5/2}$, saturated at the electron mean free path following \citet{Cowie1977}. Conduction is therefore a separate physical process from the resolution-scale conductivity included in $\Lambda_{\rm diss}$ and is not used as a numerical dissipation switch. We do not include physical viscosity or explicit Ohmic dissipation besides what is used in the shock capturing scheme. Although the magnetic field formally suppresses cross-field conduction, anisotropic Braginskii-type conduction is significantly more expensive in SPMHD and was shown in the appendix of \citet{Steinwandel2022_cluster} to give cluster-scale results that are largely consistent with the isotropic Spitzer prescription at this resolution; we therefore use the isotropic version here for computational efficiency.

The initial conditions are seeded at $z = 69$ with a spatially constant magnetic field directed along the $x$-axis, of strength $10^{-14}$~G in comoving units, corresponding to approximately $10^{-8}$~G in physical units at the seeding redshift. We acknowledge that a uniform field aligned with one Cartesian axis is not ideal: it singles out a preferred direction that has no physical motivation, and the resulting initial $B$-field topology is not what would be expected from realistic seeding mechanisms. We adopt it nonetheless because it is the simplest IC consistent with $\nabla\cdot\mathbf{B} = 0$ at machine precision, and because it has been used as the standard seed topology in essentially all previous cosmological simulations of cluster magnetic-field amplification at this resolution \citep[e.g.][]{Dolag1999, Vazza2018, Steinwandel2022_cluster}, which keeps the present results directly comparable to that body of work. The small-scale dynamo is expected to erase memory of the initial direction within a few eddy turnover times once the cluster collapses, so the orientation choice should not affect the saturated-field structure inside the cluster. As discussed in Section~\ref{sec:intro}, the elevated seed amplitude --- several orders of magnitude above the canonical Biermann-battery level of $\sim 10^{-20}$~G --- is a deliberate choice that stands in for the magnetic enrichment of the ICM by galactic outflows that we do not model explicitly, and ensures the field saturates at the observed $\sim\mu$G level in the cluster centre by $z = 0$. We performed five simulations of the same cluster, each isolating one specific choice in the cleaning solver. The calculations are summarised in the following list; for each we state both what is included and what is deliberately switched off, since the comparison between simulations is the point.

\begin{itemize}
    \item NIFTY-noclean: \emph{reference run with no Dedner-type cleaning}. The induction equation is solved in the divergence-advecting form (Eq.~\ref{eq:induction}), equivalent in SPMHD to a Powell-type scheme: monopole errors are transported with the flow but not actively removed. \emph{Does not} carry a $\psi$ field, hyperbolic propagation of the divergence error, parabolic damping, advection-like source term, or entropy back-reaction.

    \item NIFTY-clean: \emph{fiducial run with constrained Dedner-type cleaning}. Cleaning speed $c_h = \sqrt{v_A^2 + c_s^2}$ (Eq.~\ref{eq:clean_speed}), parabolic damping with $\sigma = 1$, overclean factor $\zeta = 1$; $\psi/c_h$ is evolved (Eq.~\ref{eq:evolution_grad_psi}). \emph{Includes} the advection-like source term $-\tfrac{1}{2}\psi(\nabla\cdot\mathbf{v})$ in the $\psi/c_h$ evolution equation, \emph{and} feeds the dissipated cleaning-field energy back into the entropy equation as heat. This is therefore the most complete, thermodynamically consistent realisation of the constrained Dedner scheme of \citet{Tricco2016}.

    \item NIFTY-clean2: \emph{as NIFTY-clean but with overclean factor $\zeta = 2$}, so that the cleaning timestep is allowed to follow twice the local fast magnetosonic speed. All other settings are identical to NIFTY-clean. \emph{Includes} both the advection-like source term and the entropy back-reaction.

    \item NIFTY-advect: \emph{as NIFTY-clean but with the $-\tfrac{1}{2}\psi(\nabla\cdot\mathbf{v})$ term switched off} in the $\psi/c_h$ equation. This is the source term required for strict energy conservation in the constrained derivation (Section~\ref{sec:cleaning}); this run isolates its impact in a cosmological setting by removing it from the otherwise complete scheme. \emph{Still includes} the entropy back-reaction.

    \item NIFTY-heat: \emph{as NIFTY-clean but with the parabolic-damping energy \underline{not} fed back into the entropy equation as heat}, in contrast with eq.~25 of \citet{Tricco2016}. This run isolates the impact of the entropy back-reaction; the cleaning is still stabilised by removing energy from $\psi$, but that energy is now discarded rather than deposited as heat, so the scheme is no longer thermodynamically consistent at the system level. \emph{Still includes} the advection-like source term.
\end{itemize}

We acknowledge that the run names are not perfectly self-describing. They label the term that is \emph{switched off} relative to the fiducial NIFTY-clean scheme: NIFTY-advect refers to the velocity-divergence source term $-\tfrac{1}{2}\psi(\nabla\cdot\mathbf{v})$ in the $\psi/c_h$ equation (sometimes loosely called the ``advection term'' in the SPMHD literature, and not to be confused with the bulk advection of the divergence error itself), which is dropped in that run; NIFTY-heat refers to the entropy back-reaction that converts the cleaning-field energy to thermal energy, which is dropped in that run. We retain these labels for consistency with our internal bookkeeping but make their physical meaning explicit above.


\section{Results}
\label{sec:results}

\begin{figure*}
    \centering
    \includegraphics[width=0.99\textwidth]{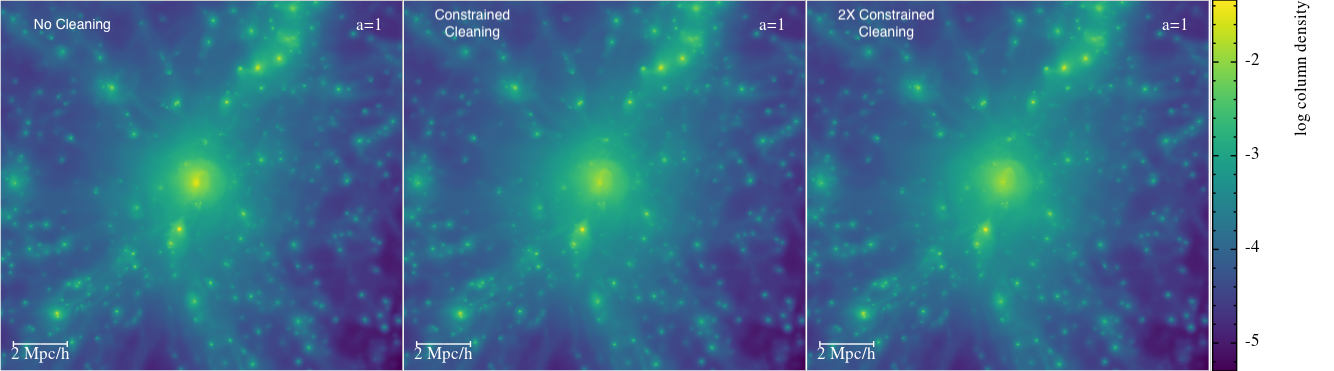}
    \includegraphics[width=0.99\textwidth]{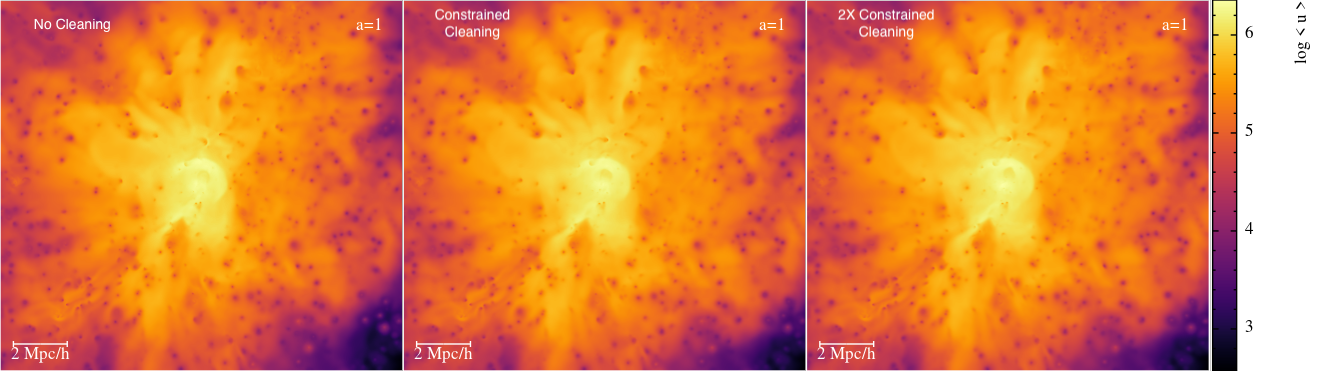}
    \caption{\textit{Top row:} Column density projections for our cluster for three of our runs, NIFTY-noclean (left), NIFTY-clean (centre) and NIFTY-clean2 (right). The central density in the run NIFTY-noclean is higher than in the runs NIFTY-clean and NIFTY-clean2. \textit{Bottom row:} Internal energy of the different clusters, the thermal structure is remains very similar in the three runs.}
    \label{fig:density}
\end{figure*}

\begin{figure*}
    \centering
    \includegraphics[width=0.99\textwidth]{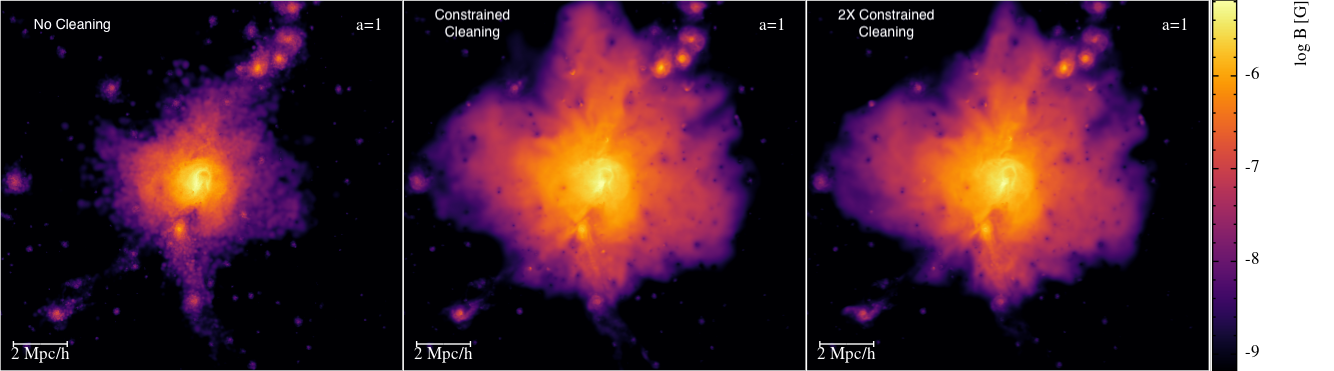}
    \includegraphics[width=0.99\textwidth]{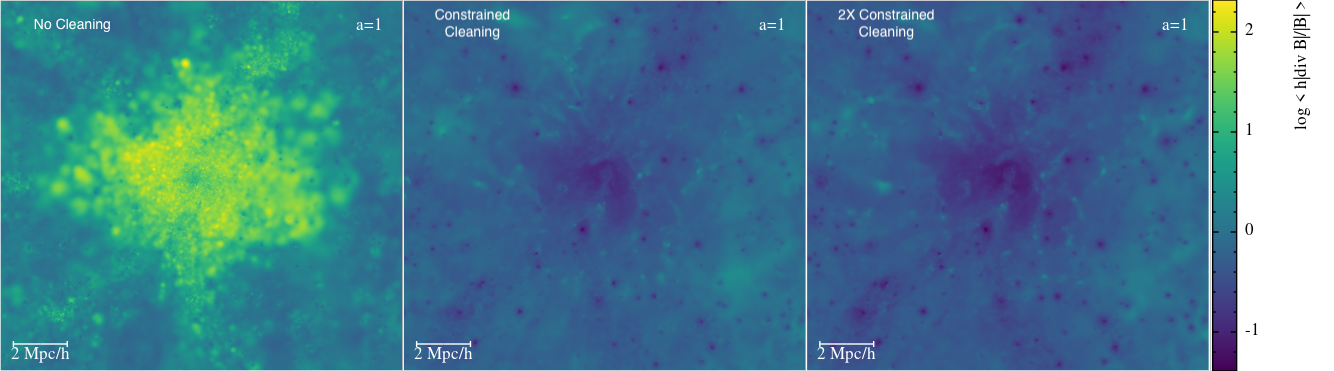}
    \caption{\textit{Top row:} Magnetic field strength for the simulations NIFTY-noclean (left), NIFTY-clean (centre) and NIFTY-clean2 (right). The magnetic field strength in the runs with active cleaning is more spread out than in the run without cleaning. We will discuss the reason for this in more detail in Sec.~\ref{sec:discussion}. \textit{Bottom row:} Normalized divergence within the cluster for the three models. We finds divergence errors reduced by up to three orders of magnitude in the central regions of the cluster in the NIFTY-clean and NIFTY-clean2 calculations.}
    \label{fig:bfld}
\end{figure*}

\begin{figure*}
    \centering
    \includegraphics[width=0.49\textwidth]{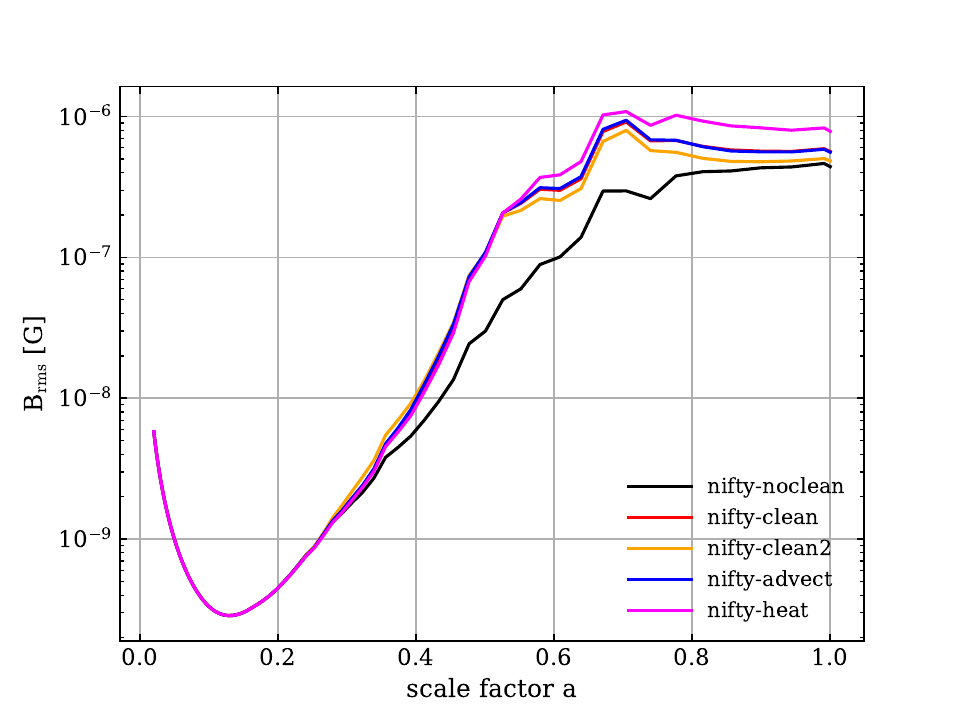}
    \includegraphics[width=0.49\textwidth]{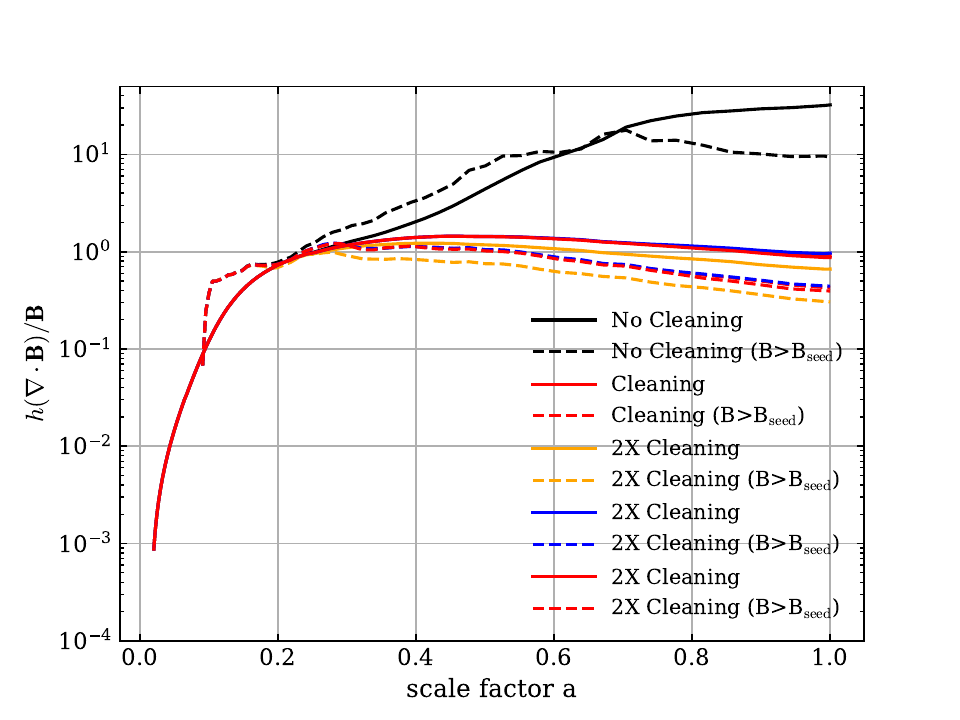}
    \caption{\textit{Left:} Mean magnetic field strength within the virial radius of the NIFTY cluster as a function of scale factor. \textit{Right:} Mean relative divergence error as a function of scale factor, for the same cluster. In both panels, line colour encodes the run: black is NIFTY-noclean (no Dedner-type cleaning), red is NIFTY-clean (fiducial cleaning, $\zeta = 1$), and gold is NIFTY-clean2 (cleaning at twice the fastest wave speed, $\zeta = 2$). Solid lines show the average taken over all gas particles; dashed lines show the average restricted to particles whose field strength exceeds the initial physical seed value of $10^{-8}$~G, i.e.\ regions where the field has been amplified above the seed level.}
    \label{fig:mean_divb}
\end{figure*}

\begin{figure*}
    \centering
    \includegraphics[width=0.49\textwidth]{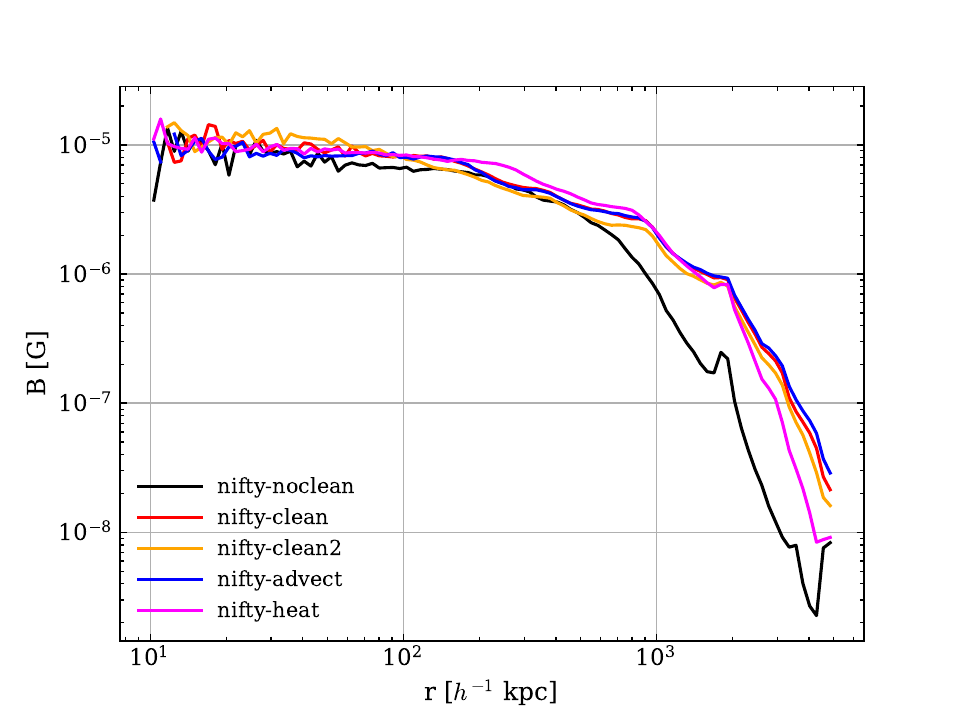}
    \includegraphics[width=0.49\textwidth]{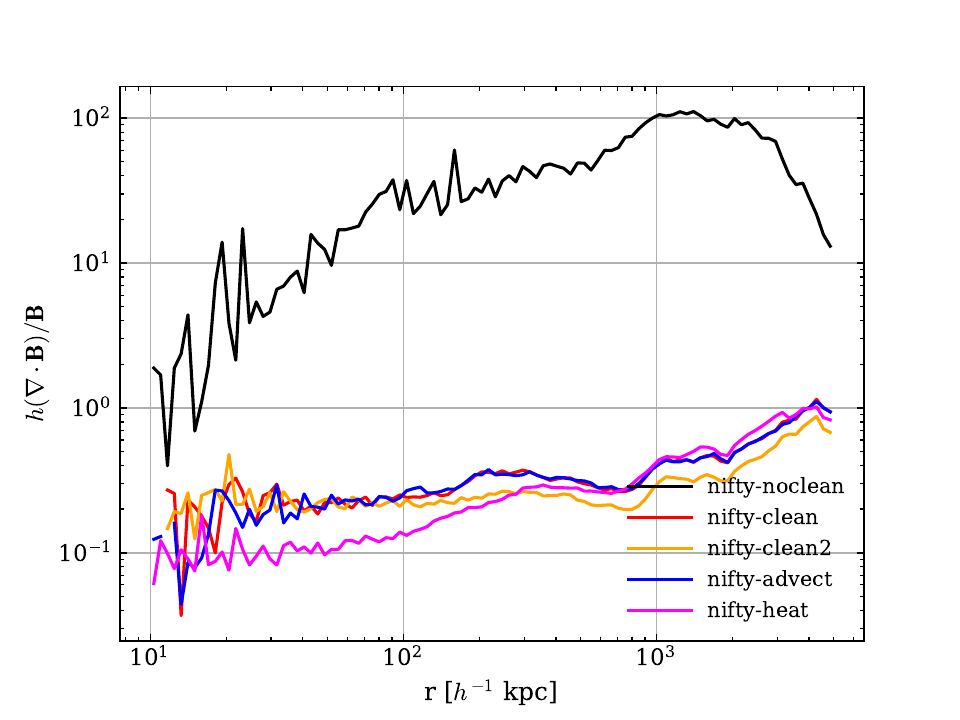}
    \caption{Radial profiles of the magnetic field strength (left) and of the relative divergence error (right) for the NIFTY cluster, for the runs without cleaning (black), with the new fiducial constrained cleaning scheme (red), and with $2\times$ faster cleaning than the fastest wave speed (gold). The radial entropy profile is shown separately in Fig.~\ref{fig:entropy}.
    }
    \label{fig:radial_nifty}
\end{figure*}

\begin{figure}
    \centering
    \includegraphics[width=0.49\textwidth]{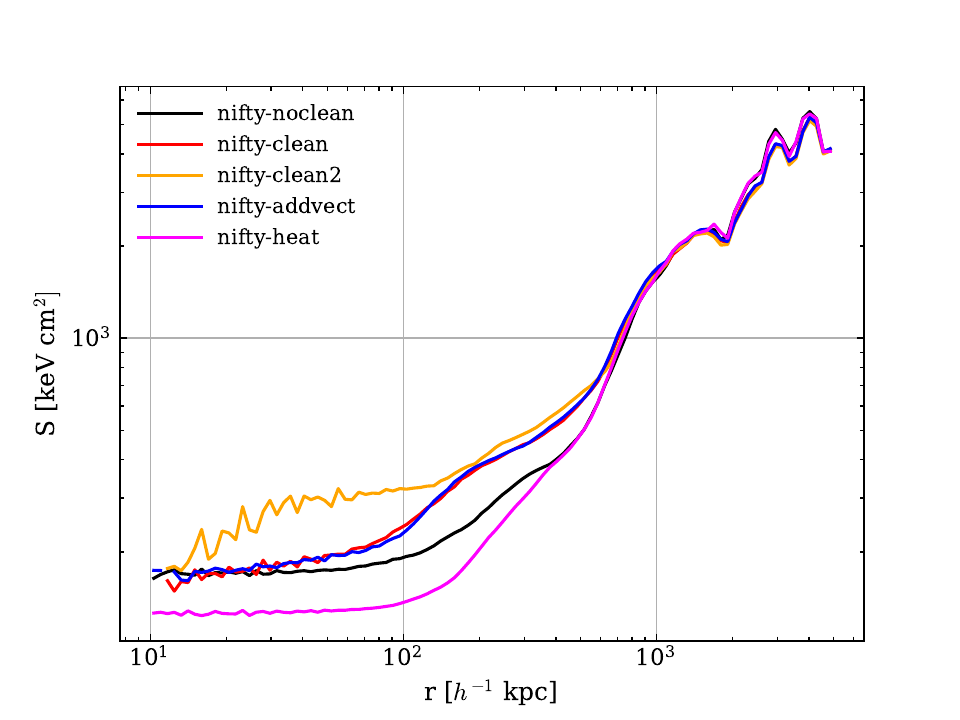}
    \caption{Radial entropy profiles of our simulated clusters. The entropy profile is quite sensitive to the adopted cleaning method, which is important to point out. The entropy core in the cluster can actually be stabilized with higher cleaning speed, when we convert the energy stored in the cleaning field into heat, i.e. following eq.~21 of \citet{Tricco2016}. If we exclude this term the entropy core settles at a lower value. Removing the advection term from the cleaning scheme has a rather little impact on the entropy profile.}
    \label{fig:entropy}
\end{figure}

Fig.~\ref{fig:density} shows column-density projections (top row; $\int \rho {\rm d}z$) and column-averaged internal energy ($<u> \equiv {\int \rho u {\rm d}z}\,/\,{\int \rho {\rm d}z}$; bottom row), for the models NIFTY-noclean, NIFTY-clean and NIFTY-clean2 (left to right).

We find only minor differences in the global evolution of these quantities due to the inclusion of the new cleaning scheme. However, the central density in the run NIFTY-noclean is slightly higher by roughly a factor of two than in the runs NIFTY-clean and NIFTY-clean2, the origin of which we will discuss in greater detail in Sec.~\ref{sec:discussion}. For now it is sufficient to note that this effect is related to whether or not we dissipate the energy stored in the cleaning field as heat.

Fig.~\ref{fig:bfld} shows a slice of the magnetic field strength (top row) and the column-averaged relative divergence of the field (bottom row), again for the three runs NIFTY-noclean, NIFTY-clean and NIFTY-clean2 (left to right). The magnetic field in the runs with the new constrained cleaning of \citet{Tricco2016} is more spread out. Again, some of this originates from how we chose to dissipate the energy stored in the cleaning field --- we discuss the details in Section~\ref{sec:discussion}. The bottom row reveals that the new scheme outperforms {\sc Gadget}'s older Powell divergence-preservation scheme by at least 2.5 orders of magnitude. Increasing the overclean factor to $\zeta = 2$ in the run NIFTY-clean2 reduces the divergence error further than the fiducial $\zeta = 1$ scheme, but the improvement in the cluster is more modest than in the idealised advection test (Fig.~\ref{fig:advection_time}, where $\zeta = 2$ is roughly an order of magnitude better than $\zeta = 1$). In the cluster, the additional reduction is a factor of 2--3 compared to NIFTY-clean (right panel of Fig.~\ref{fig:radial_nifty}); the smaller relative gain reflects the continuous generation of divergence errors that occurs in cosmological simulations, which was not present in the test problem where $\nabla\cdot{\bf B} \neq 0$ is introduced only in the initial conditions. All three simulations have a similar runtime down to the 2 per cent level, i.e.\ the improved cleaning comes essentially for free. We note that pushing $\zeta$ beyond a factor of four does begin to slow the code dramatically and that the gain in divergence control is small.

In Fig.~\ref{fig:mean_divb} we show the mean magnetic field strength within the virial radius of the simulated cluster as a function of scale factor (left panel) and the evolution of the relative divergence as function of scale factor (right panel). Interestingly, we find that with divergence cleaning the field grows at a slightly steeper rate than without the divergence cleaning prescription. We attribute this to excess dissipation of magnetic energy in simulations with high divergence errors due to unphysical small-scale noise in the magnetic field, an effect also seen in turbulence simulations in SPH where insufficient shock viscosity can lead to excess  dissipation of energy \citep{PriceFederrath2010}. \citet{Tricco2012} also showed that divergence cleaning could produce much better control of divergence errors without excessively dissipating magnetic energy, compared to using resolution-scale resistivity alone. Hence divergence cleaning leads to an effectively higher magnetic Prandtl number.

It is clear from the right hand panel of Fig.~\ref{fig:mean_divb} that the mean divergence error as function of scale factor is reduced by at least one order of magnitude. We also measure the relative divergence error in the regions that exceed the initial seed field to quantify the divergence error in the regions where the magnetic field is strongly amplified, which we plot as the dashed lines for each run. It is interesting to note that the divergence is generally speaking lower in the regime where the magnetic field is strong, even in the case where there is only the Powell cleaning scheme applied where we find that the divergence is lower in the highly magnetized regions at the end of the simulations by roughly a factor of four. While this trend is somewhat present in the two runs NIFTY-clean and NIFTY-clean2 it is weaker and the divergence in the magnetized regions is only factor of two lower than in the whole simulation region. The advection term in the cleaning equations as well as the exact way we choose to dissipate the energy stored in the cleaning equations does not have a strong effect on the outcome of any of these quantities, manifesting in the fact that the runs NIFTY-advect and NIFTY-heat do not change the relative divergence error very strongly. However, they do affect the saturation value of the field down to redshift zero. In particular, the run NIFTY-heat --- in which we convert the energy stored in the cleaning field back into heat --- shows higher central densities and correspondingly lower entropy in the core, so that the magnetic field is more strongly adiabatically compressed and its saturation value is increased compared to the other cleaning runs.

In Fig.~\ref{fig:radial_nifty} we show the radial profiles of the magnetic field strength on the left as well as the radial profiles of the relative divergence of the magnetic field. As expected from investigation of the first row of Fig.~\ref{fig:bfld} we find that in the simulations with the new active cleaning scheme following the prescription of \citet{Tricco2016}, the magnetic field strength out larger radii is higher by roughly half a dex compared to the run with active Powell-cleaning NIFTY-noclean. It is interesting to note that the magnetic field strength in the run NIFTY-heat drops quicker than in the other runs with the new cleaning scheme. In NIFTY-heat the energy stored in the cleaning field $\psi$ is removed by the parabolic damping but \emph{not} returned to the gas as heat, so the cleaning channel acts as a pure sink of magnetic energy and depresses $|\mathbf{B}|$ at larger radii relative to the runs in which this energy is recycled into the entropy equation. However, the magnetic field strength in the radial profile is still higher in all the runs in which we include the new constrained cleaning prescription. The radial profile of the relative divergence reveals that the divergence error is at least one order of magnitude lower in all runs compared to the run NIFTY-noclean.

Fig.~\ref{fig:entropy} shows the radial entropy profiles for all five runs that we carried out. The entropy profiles of the cluster turns out to be most sensitive to the the numerical details of the cleaning scheme that is adopted. For instance, the run NIFTY-heat has the lowest entropy in the cluster centre. This is consistent with the highest magnetic fields at saturation in this run in the time-evolution in the left hand panel of Fig.~\ref{fig:mean_divb}. The reason that the entropy is the lowest in NIFTY-heat is relate to the fact that this run does not have an additional source term for the entropy while in all the other runs we dissipate the energy stored in the cleaning field as heat back into the systems. Hence, this is literally a source of entropy where the cleaning speed is high, i.e. in the center of the cluster. The reason that the entropy is larger in the run NIFTY-clean2 in the central regions is related of the fact that the cleaning is faster by a factor of two, dissipating the energy more effectively. The outskirts of the cluster show convergence in the entropy profile, independent of the details of the adopted cleaning scheme. We note that there is no consensus to this day on what the exact shape of the cluster entropy profile should be, but it is apparent that the cleaning scheme can affect this, specifically if we choose to add the energy stored in the cleaning field $\psi$ as an additional source term for the entropy. Historically, SPH struggles with cored entropy profiles in non-radiative simulations of galaxy cluster formation and thus it is interesting to point out that with active cleaning we find a tendency to realize a cored cluster profile.

\section{Discussion} \label{sec:discussion}

The headline numerical result of this work is straightforward: the constrained Dedner-type cleaning scheme of \citet{Tricco2012} and \citet{Tricco2016}, implemented here in {\sc OpenGadget3}, reduces the relative divergence error in our cluster volume by 2--3 orders of magnitude compared to the Powell-type advecting scheme already used in {\sc Gadget}, at no measurable runtime cost. By the metric on which a divergence-cleaning scheme has to be judged --- how much it suppresses $\nabla\cdot\mathbf{B}$ --- the implementation works as intended, and the gain is uniform across both the cluster core and the outskirts (Fig.~\ref{fig:radial_nifty}, right panel; Fig.~\ref{fig:mean_divb}, right panel).

\subsection{Sensitivity to numerical choices in the cleaning solver}

Beyond the headline result, the four cleaning runs (NIFTY-clean, NIFTY-clean2, NIFTY-advect, NIFTY-heat) isolate the impact of three distinct numerical choices that are available within the constrained Dedner-type framework.

\paragraph*{Overclean factor.} Increasing $\zeta$ from $1$ to $2$ (NIFTY-clean2) gives a further factor of 2--3 reduction in the divergence error throughout the cluster volume, smaller than the order-of-magnitude gain seen in the idealised advection test (Fig.~\ref{fig:advection_time}). The smaller relative gain in the cluster reflects the fact that the divergence floor in a cosmological setting is set {\bf by the balance between generation of divergence errors in} accretion shocks and the strong dynamic range of $\rho$ and $\mathbf{v}$ {\bf and the speed with which the cleaning can remove them}.

\paragraph*{Velocity-divergence source term.} Switching \emph{off} the $-\tfrac{1}{2}\psi(\nabla\cdot\mathbf{v})$ term in the $\psi/c_h$ equation (NIFTY-advect) makes essentially no difference at the level of the radial profiles of $|\mathbf{B}|$, $\nabla\cdot\mathbf{B}$, and entropy: NIFTY-advect tracks NIFTY-clean to within the noise of the runs. This confirms in a cosmological cluster regime --- with virial velocities $\sim 10^3$~km~s$^{-1}$, strong central accretion, and merger-driven shocks --- the prediction made by \citet{Tricco2012} and \citet{Tricco2016} from idealised tests, namely that this term is subdominant to the hyperbolic and parabolic cleaning channels. It is reassuring that the dynamic range of velocity divergence in a cluster does not invalidate that conclusion.

\paragraph*{Parabolic-damping back-reaction.} The largest impact among the three numerical choices comes from whether the energy removed from the cleaning field by the parabolic damping is fed back into the entropy equation as heat (NIFTY-clean, NIFTY-clean2, NIFTY-advect, following eq.~25 of \citealt{Tricco2016}) or simply discarded (NIFTY-heat). Switching the back-reaction \emph{off} lowers the central entropy and raises the saturated central magnetic field, because without the deposited heat the core stays colder, contracts further, and adiabatically compresses the field. Mechanistically, the back-reaction is what makes the parabolic damping thermodynamically consistent at the level of the total energy budget: the energy taken out of $\psi$ has to go somewhere, and feeding it into the entropy is the second-law-consistent choice. In the absence of this back-reaction (NIFTY-heat), the cleaning field acts as a pure sink that simply removes energy from the system. Both options are numerically valid --- they correspond to two different statements about what one is willing to discard --- but the heating-on choice is the one that closes the energy budget. We therefore consider NIFTY-clean (back-reaction on) the more physically defensible default for cosmological MHD applications, even though the absolute thermodynamic shift introduced by switching the back-reaction off is modest. Our results provide a quantitative estimate of how large that shift is: a small but measurable reduction in central entropy and a corresponding increase in the saturated magnetic field strength when the back-reaction is switched off.

\subsection{Physical interpretation of the outskirts amplification}

The most striking physical difference between the cleaning and no-cleaning runs is in the cluster outskirts ($r \approx 1$--$3~h^{-1}$~Mpc), where the magnetic field strength is amplified by a factor of 5--10 in NIFTY-clean and NIFTY-clean2 relative to NIFTY-noclean (Fig.~\ref{fig:radial_nifty}, left panel). The cluster core, in contrast, is essentially unaffected. We argue that this asymmetry is physically meaningful, not numerical:
\begin{itemize}
    \item Uncontrolled $\nabla\cdot\mathbf{B}$ errors source an unphysical Lorentz force in the momentum equation that scales as $\propto \nabla\cdot\mathbf{B}$ and is partially subtracted in the Powell/B{\o}rve form, but not eliminated. The residual force is a numerical noise term in the velocity-magnetic-field coupling that drives the small-scale dynamo.
    \item In the cluster core the magnetic field is dynamo-saturated, so the \emph{relative} error $|\nabla\cdot\mathbf{B}|\,h / |\mathbf{B}|$ is small and the residual force is dynamically negligible. In the outskirts the field is several orders of magnitude weaker, so the same absolute $\nabla\cdot\mathbf{B}$ error becomes fractionally large, and the unphysical force injects noise into the dynamo at exactly the level that is most damaging --- where the dynamo is still in its growing, sub-saturated phase.
    \item Removing this noise, by reducing $\nabla\cdot\mathbf{B}$ throughout the volume, lets the small-scale dynamo amplify the field uninhibited in the outer regions, giving the factor-5--10 increase we observe. The cleaned runs therefore represent a more faithful realisation of the dynamo physics, not an artefact of the cleaning step itself.
    \item Crucially, the divergence error is reduced \emph{at all radii} (Fig.~\ref{fig:radial_nifty}, right panel), not just in the centre. The outskirts amplification is therefore not a redistribution of $\nabla\cdot\mathbf{B}$ from the core to the outskirts, which would be the natural alternative explanation; the cleaning genuinely removes divergence rather than relocating it.
\end{itemize}

We acknowledge a resolution caveat: at $r \gtrsim 1.5$~Mpc the gas resolution is approximately $10^8\,h^{-1}$~M$_\odot$ per particle, which is the regime in which uncleaned-divergence errors are also expected to be largest. The factor-5--10 amplification is therefore the regime where the choice of divergence-control scheme matters most. That this is also the regime of poorest resolution does not invalidate the result, but we emphasise that a quantitatively converged outskirts magnetic-field strength would require resolution studies of the kind reported in \citet{Steinwandel2022_cluster}, which lie outside the scope of the present paper.

\subsection{Recommendations for cosmological SPMHD}

Based on the above we recommend three settings for cosmological SPMHD applications that use the constrained Dedner-type cleaning of \citet{Tricco2012, Tricco2016}: (i) include the cleaning by default --- the divergence-error reduction of 2--3 orders of magnitude comes essentially for free; (ii) use the natural cleaning speed $c_h = \sqrt{v_A^2 + c_s^2}$ with overclean factor $\zeta = 1$, since pushing $\zeta$ harder gives diminishing returns in a cosmological setting; (iii) keep the parabolic-energy back-reaction on, since this closes the energy budget and is the second-law-consistent choice. The $-\tfrac{1}{2}\psi(\nabla\cdot\mathbf{v})$ source term may safely be omitted if a slightly simpler implementation is desired, although there is no strong reason to drop it.

\section{Conclusions}
\label{sec:conclusions}
We have demonstrated the following key results:
\begin{enumerate}
\item Implementing the constrained, energy-conserving Dedner-type cleaning algorithm of \citet{Tricco2012} and \citet{Tricco2016} in {\sc OpenGadget3} leads to a dramatic reduction in divergence errors in cosmological simulations of galaxy clusters. Specifically, divergence errors are reduced by 2--3 orders of magnitude relative to the previously employed Powell-only SPMHD scheme in {\sc Gadget}, which merely preserves the divergence without actively removing it. This improvement ensures that the magnetic field remains closer to the physically required solenoidal condition, enhancing the overall reliability of the simulation results.

\item Large divergence errors, as observed in simulations that rely solely on the Powell scheme, have a systematic impact on the properties of the magnetic field, particularly in the outer, poorly resolved regions of clusters. In these regions, the growth of magnetic energy through dynamo amplification is suppressed, and the magnetic field strengths are generally underestimated. By actively controlling divergence, the constrained cleaning scheme allows for more accurate representation of field amplification processes, capturing stronger magnetic fields and more realistic dynamo behavior in low-density, under-resolved cluster outskirts.

\item The cleaning scheme also improves the qualitative appearance of the magnetic field. Compared to simulations without cleaning, the resulting field is overall smoother and less patchy, with coherent structures extending over larger regions. This smoother field distribution not only better reflects the expected physical behavior of magnetised plasma in galaxy clusters but also helps reduce numerical artifacts that can arise from irregular, spurious divergence in under-resolved regions.

\item Even in simulations with comparatively low resolution, the inclusion of divergence cleaning leads to magnetic field geometries that are in closer agreement with those obtained in higher-resolution simulations. This consistency indicates that the cleaning scheme not only reduces numerical errors but also stabilizes the evolution of the magnetic field across different resolutions, improving the physical fidelity and robustness of the results. As a result, conclusions drawn about the magnetic structure of galaxy clusters become more reliable, even when computational limitations prevent extremely high resolution.

\end{enumerate}
Furthermore, the new energy-conserving cleaning scheme introduces no additional computational overhead and is now publicly available within the {\sc OpenGadget3} code, providing the community with a practical and efficient tool for conducting accurate magnetohydrodynamical simulations in a cosmological context.

\section*{Acknowledgements}

We thank the anonymous referee for a detailed report that greatly improved the quality of the paper. UPS and DJP thank the organizers of the Flatiron Center for Computational Astrophysics 2023 fluid dynamics summer school for inviting them both to teach, during which this project was initiated. The simulation code and the simulations in this work have been carried out on the Flatiron in-house cluster ``rusty''. UPS thanks Ludwig B\"oss, Klaus Dolag, R\"udiger Pakmor and Volker Springel for insightful comments. DJP thanks N. Cuello, F. M\'enard and everyone in Grenoble for hosting him on sabbatical where this paper was completed.

\section*{Data Availability}

The data in this paper will be made available based on reasonable request to the corresponding author.



\bibliographystyle{mnras}
\bibliography{example} 




\appendix

\section{Idealised Tests}
\label{sec:tests}

\begin{figure*}
    \centering
    \includegraphics[width=\textwidth]{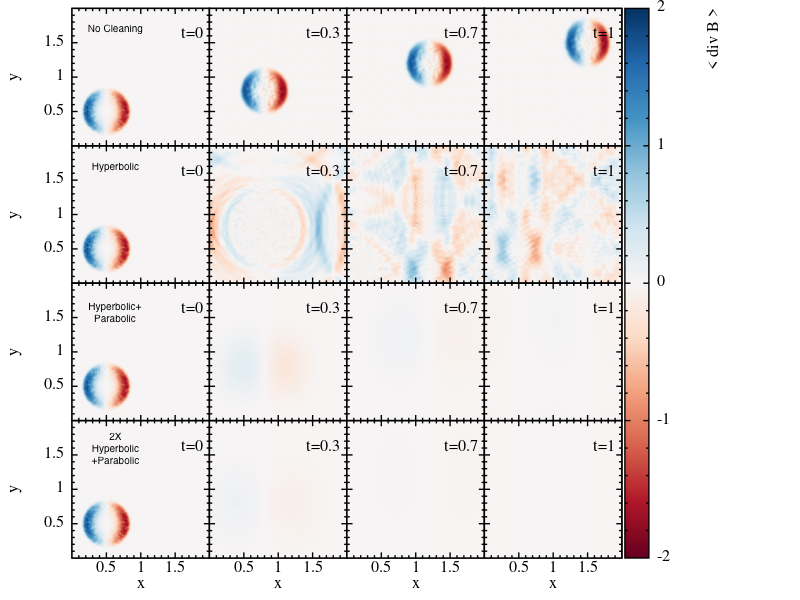}
    \caption{Divergence advection test of an artificially introduced divergence error in the systems that is advected alongside the diagonal. We present four simulation runs with no cleaning (top row), hyperbolic cleaning only (second row), hyperbolic and parabolic cleaning (third row, fiducial model) as well as a fourth run with parabolic and hyperbolic cleaning that allows for a faster cleaning speed than the fastest wave speed by a factor of two.
    }
    \label{fig:advection}
\end{figure*}

\begin{figure}
    \centering
    \includegraphics[width=\columnwidth]{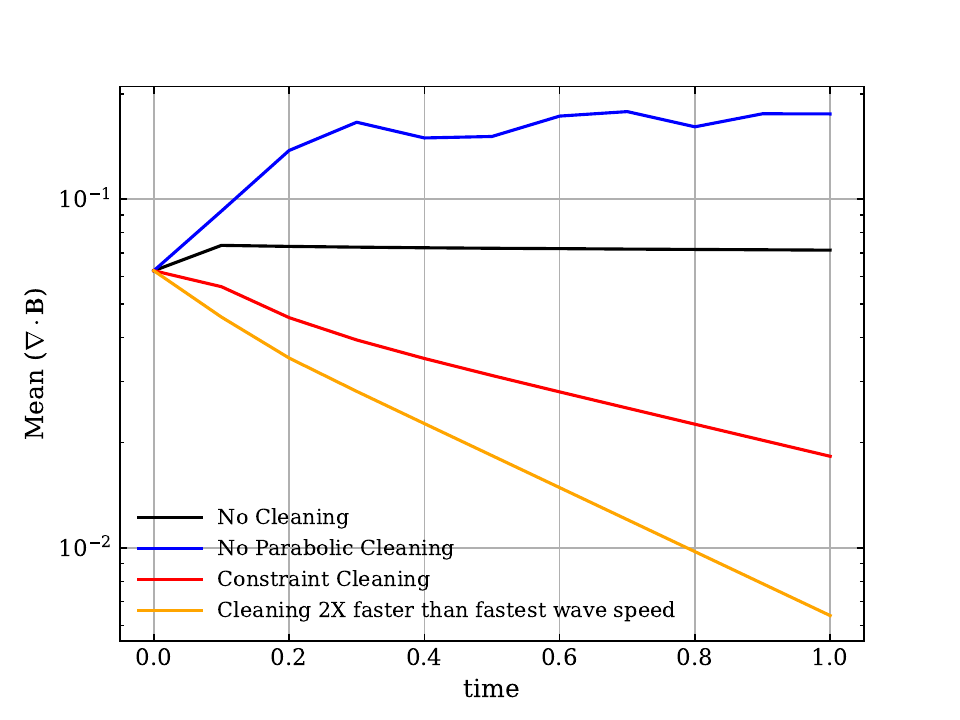}
    \caption{We show the time evolution of the mean value of the divergence of the magnetic field for the four advection tests carried out. In the absence of cleaning, we find that the code is preserving the initial divergence error (black line). For hyperbolic cleaning only we find that the code is increasing the mean value of the divergence compared to the reference run without cleaning but stabilizes at a higher value by around a factor of 2. For the fiducial cleaning scheme, we find that the divergence error is reduced by the end of the simulation by around a factor of 5 (red line). If we allow for cleaning speeds that are 2 times faster than the fastest wave speed we find an improvement by a factor of 10 at t=1.}
    \label{fig:advection_time}
\end{figure}

\begin{figure*}
    \centering
    \includegraphics[width=0.95\textwidth]{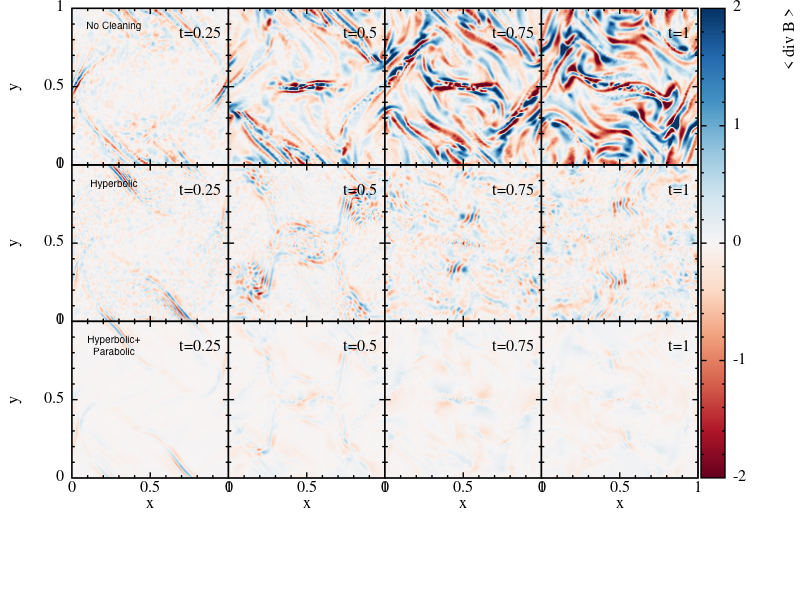}
    \vspace{-2cm}
    \caption{Divergence advection test of the Orszag-Tang vortex' divergence error. We present four simulation runs with no cleaning (top row), hyperbolic cleaning only (second row), hyperbolic and parabolic cleaning (third row, fiducial model) as well as a fourth run with parabolic and hyperbolic cleaning that allows for a faster cleaning speed than the fastest wave speed by a factor of two.}
    \label{fig:orszag_tang}
\end{figure*}

\begin{figure}
    \centering
    \includegraphics[width=0.49\textwidth]{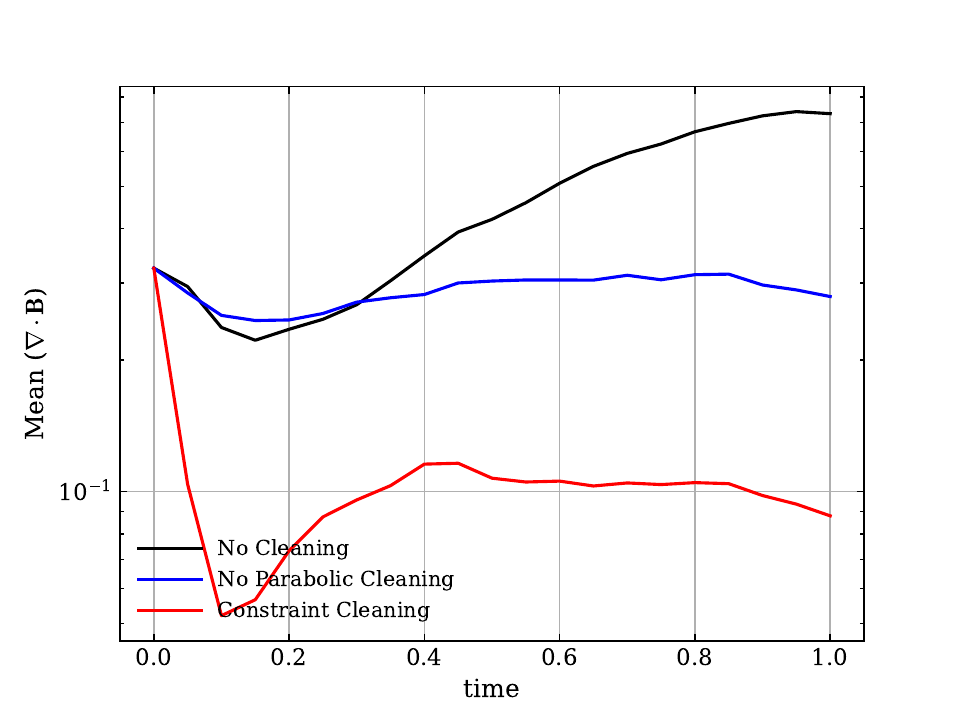}
    \caption{We show the time evolution of the mean value of the divergence of the magnetic field for the four advection tests carried out. In the absence of cleaning, we find that the code is preserving the initial divergence error (black line). For hyperbolic cleaning only we find that the code is increasing the mean value of the divergence compared to the reference run without cleaning but stabilizes at a higher value by around a factor of 2. For the fiducial cleaning scheme, we find that the divergence error is reduced by the end of the simulation by around a factor of 5 (red line). If we allow for cleaning speeds that are 2 times faster than the fastest wave speed we find an improvement by a factor of 10 at t=1.}
    \label{fig:orszag_tang_time}
\end{figure}

\begin{figure*}
    \centering
    \includegraphics[width=\textwidth]{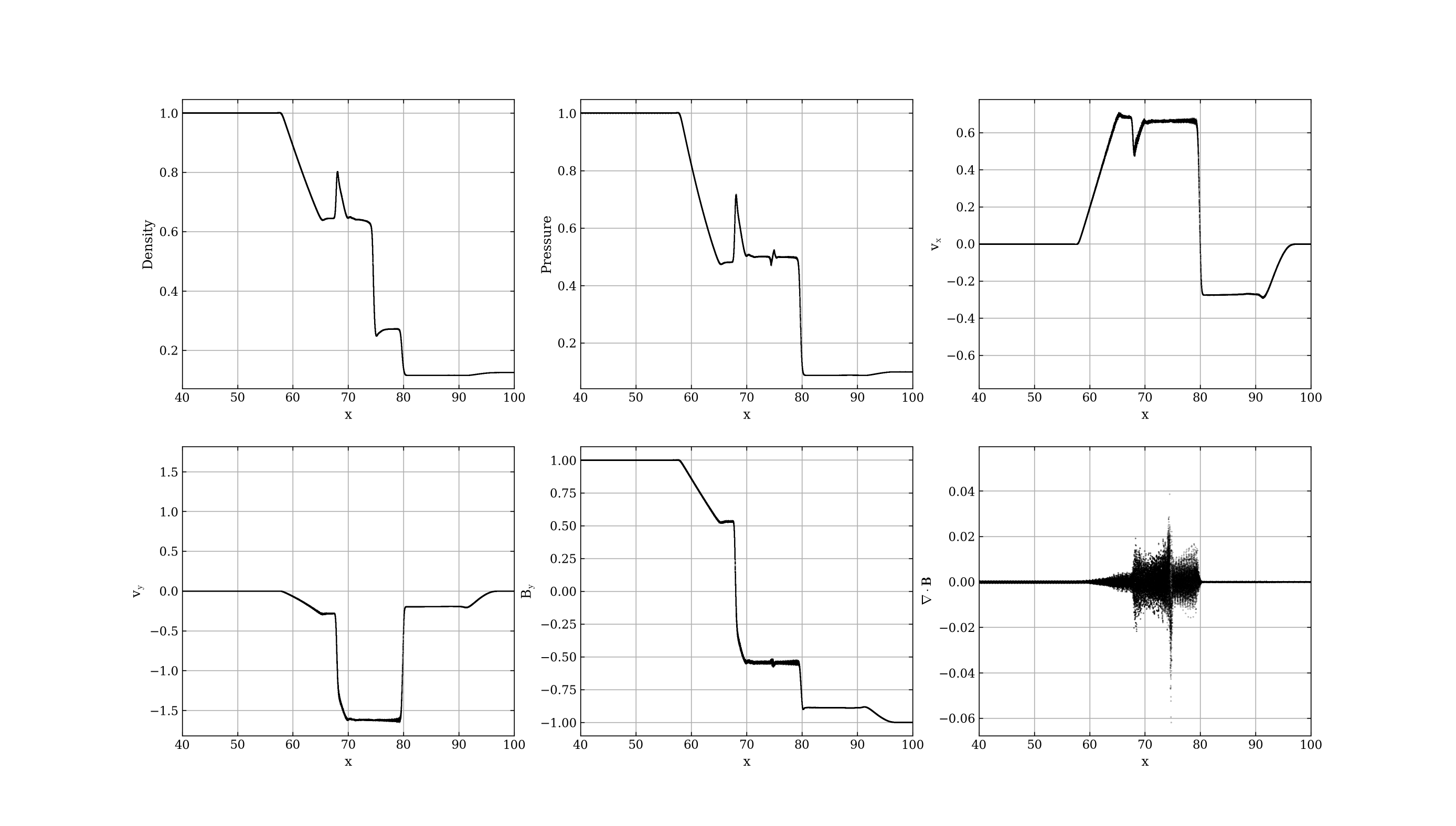}
    \caption{Simulation of a Brio-Wu shock tube at t=7. We show the density (top left), the pressure (top center), the x-velocity (top right), the y-velocity (bottom left), the y-component of the magnetic field (bottom center), and the divergence of the magnetic field (bottom right).}
    \label{fig:shock_noclean}
\end{figure*}

\begin{figure*}
    \centering
    \includegraphics[width=\textwidth]{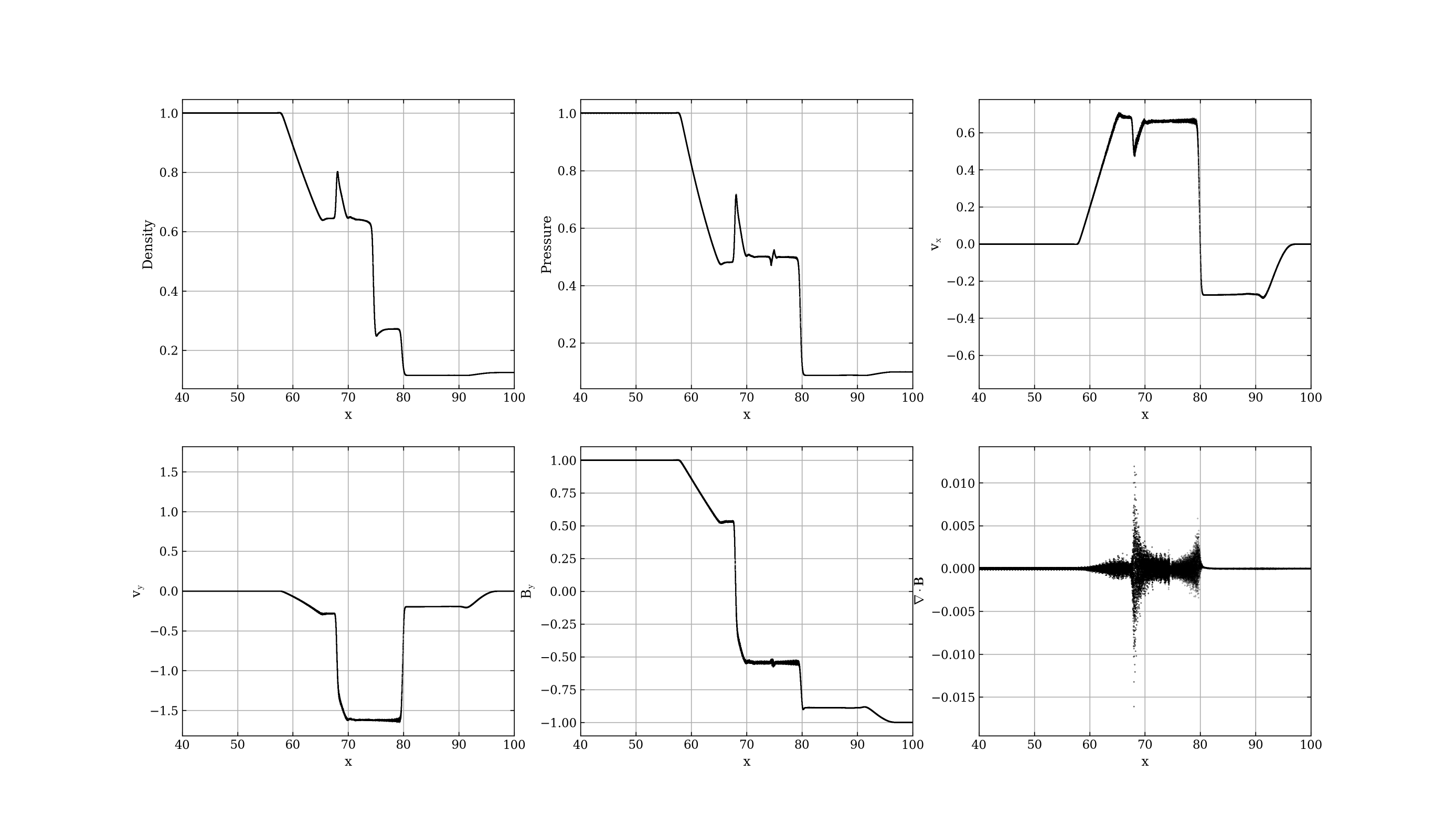}
    \caption{Same as Fig.~\ref{fig:shock_noclean} but for a simulation includes the hyperbolic and parabolic cleaning terms (fiducial model). }
    \label{fig:shock_clean}
\end{figure*}

\begin{figure}
    \centering
    \includegraphics[width=0.49\textwidth]{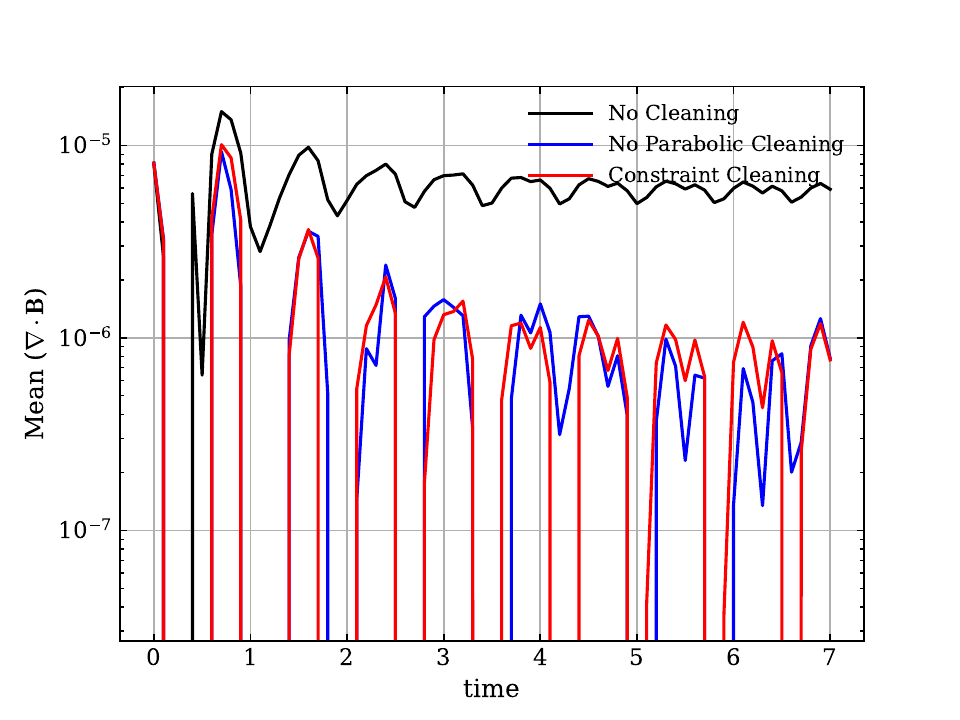}
    \caption{We show the time evolution of the mean value of the divergence of the magnetic field for the four advection tests carried out. In the absence of cleaning, we find that the code is preserving the initial divergence error (black line). For hyperbolic cleaning only we find that the code is increasing the mean value of the divergence compared to the reference run without cleaning but stabilizes at a higher value by around a factor of 2. For the fiducial cleaning scheme, we find that the divergence error is reduced by the end of the simulation by around a factor of 5 (red line). If we allow for cleaning speeds that are 2 times faster than the fastest wave speed we find an improvement by a factor of 10 at t=1.}
    \label{fig:shocktube_time}
\end{figure}

\subsection{Divergence Advection Test}

We start with the popular divergence advection test as it has been carried out by several groups to test divergence cleaning schemes in SPH. The test is initialized in a periodic domain with $0.0 < x < 2.0$, $0.0 < y < 2.0$, and $0.0 < z < 0.1$ we note that we carry the test out in three spatial dimensions where the z-direction is factor of 20 thinner than the x and y directions. The test is carried out with a Wendland C4 kernel where we do not apply the nonphysical kernel bias correction of \citet{Dehnen2012}. The density is initialzied as $\rho=1$, the pressure as $P=6$ and we adopt $\gamma=5/3$ with an equation of state as $P= \rho u (\gamma-1)$. The divergence perturbation is advected with $v_\mathrm{x} = v_\mathrm{y}$=1. We set $v_\mathrm{z}$ to zero. The magnetic field is set to $B_\mathrm{x} = B_\mathrm{y}=0 $, except for a small perturbation within the x-component of the magnetic field given via:
\begin{align}
    B_\mathrm{x} = \frac{1}{4 \pi} \left[\left(\frac{r}{r_0}\right)^8 - 2\left(\frac{r}{r_0}\right)^4 +1 \right],
\end{align}
with $r=\sqrt{x^2 + y^2}$ and $r_0 = 1/\sqrt{8}$. The z-component of the magnetic field is set to $B_\mathrm{z}=1/\sqrt{4 \pi}$. The conditions lead to a $\beta$ around 150. Since the $\beta$ is high there is not necessarily the need to correct for Tensile-instability hence we switch it off, alongside the artificial resistivity to test energy conservation of the method. These are the same settings as adopted by \citet{Tricco2016}. The particle positions are initialized on a hydrodynamical glass distribution with 512,000 particles, which is obtained by running the code with very large viscosity until equilibrium is reached. This is again different from the procedure in \citet{Tricco2016}, who simply adopted particles based on a triangular lattice.
In Fig.~\ref{fig:advection} we show the results of the test simulation to validate the basic functionality of the the method in the simulation code Gadget. The top row shows the results with no cleaning applied, the second row shows the results without parabolic (only hyperbolic) cleaning terms, and the third row shows the full scheme that implements both, the hyperbolic and parabolic cleaning terms with available cleaning speed. In the bottom row, we demonstrate a simulation in which the cleaning speed is 2 times faster than the fastest wave speed in the simulation, to demonstrate that our implemented timestep limiter for the cleaning speed is working properly. In Fig.~\ref{fig:advection_time} we show the time evolution of the divergence of the magnetic field for our 4 advection tests. This test clearly demonstrates that the divergence error is preserved when no cleaning is applied. In the case in which we only allow for hyperbolic cleaning, the situation becomes worse because there is no damping of the waves introduced by the hyperbolic term alone. However, the divergence error is reduced by about an order of magnitude when the fiducial scheme in row three is applied.

\subsection{Orszag-Tang Vortex}

The next test we perform is the Orszag-Tang-Vortex test \citet{Orszag1979}. Hereby, we closely follow the setup of \citet{Tricco2016}. Particle positions are again initialized via a hydrodynamical glass distribution with 512,000, with density $\rho = 25 / 36|pi$, pressure $P=5/12\pi$, velocity v$_\mathrm{x} = -\sin{2\pi \cdot y}$, v$_\mathrm{y} = \sin{2\pi \cdot x}$, v$_\mathrm{z} = 0$, and magnetic field B$_\mathrm{x} = -\sin{4\pi \cdot y}$, B$_\mathrm{y} = -\sin{2\pi \cdot x}$ and B$_\mathrm{z} = 0$, in 3D periodic domain, which spans $0.0 < x < 1.0$, $0.0 < y < 1.0$, and $0.0 < z < 0.1$. We show the result of the test for simulations without cleaning (top row), with hyperbolic cleaning only (center row), and with the constrained divergence cleaning (bottom row), for four different points in time at t=0.25, t=0.5, t=0.75, and t=1.0 in Fig.~\ref{fig:orszag_tang}. We show the time evolution of the divergence error for these simulations in Fig.~\ref{fig:orszag_tang_time}.

\subsection{Brio-Wu Shocktube}

Finally, we tested the scheme against a Brio-Wu shock tube \cite{Brio1988}. The test is realized within a three-dimensional domain with extent $0.0 < x < 140$, $0.0 < y < 1$, and $0.0 < z < 1$ with periodic boundary conditions. The test itself is realized with a total of 600,000 particles on a glass-like distribution. The shock tube has an initial left-hand state given via $[\rho, P, v_\mathrm{x}, v_\mathrm{y}, v_\mathrm{z}, B_\mathrm{x}, B_\mathrm{y}, B_\mathrm{z}]$ = $[1, 1, 0, 0, 0, 0.75, 1, 0]$ and an initial right hand state given via $[\rho, P, v_\mathrm{x}, v_\mathrm{y}, v_\mathrm{z}, B_\mathrm{x}, B_\mathrm{y}, B_\mathrm{z}]$ = $[0.125, 0.1, 0, 0, 0, 0.75, -1, 0]$. We performed three versions of this test case, without cleaning (Fig.~\ref{fig:shock_noclean}), with hyperbolic cleaning only, and with the fiducial scheme that is realizing the combination of hyperbolic and parabolic cleaning (Fig.~\ref{fig:shock_clean}).

\section{Cosmological Tests}

One addition in this work compared to previous implementations of the constrained cleaning scheme \citep[e.g.][]{Tricco2016, Wissing2023} is its application to cosmological simulations. We will use two popular cosmological test cases to validate that the scheme is working in co-moving units as intended.

\subsection{Magnetized Zeldovich Pancake}

\begin{figure*}
    \centering
    \includegraphics[width=0.99\textwidth]{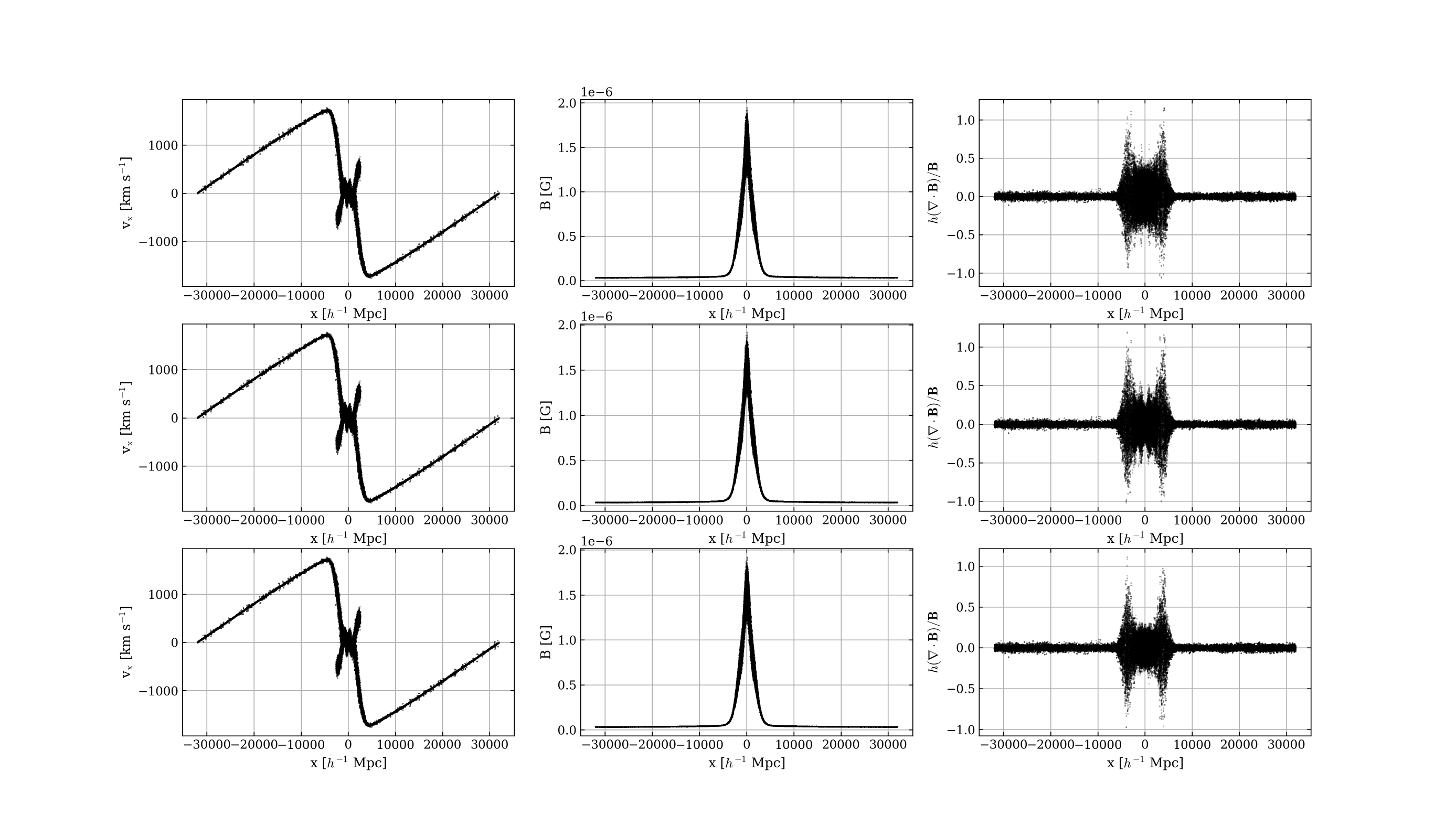}
    \caption{Results of the magnetized Zeldovich pancake test}
    \label{fig:zeldovich}
\end{figure*}

The Zeldovich-pancake is a validation test to demonstrate the basic capabilities of a code to integrate a system in a co-moving unit system. The goal of this test is less to show that the cleaning scheme is working but rather to show that it is not interfering with the expected analytic solution of the unmagnetized problem. The test has been introduced by \citet{Zeldovich1970} and is well suited to test high Mach number flows in combination with highly anisotropic particle distributions and low specific internal energies. The test is initialized at a redshift of z$_\mathrm{i} = 100$ with a single sine wave perturbation in the density field which will be tracked down to redshift z$_\mathrm{c}=1$ at which it is transitioning to the caustic formation that we follow all the way down to redshift zero. The initial conditions are as follows:
\begin{align}
    x = x_\mathrm{i} - \frac{1+z_\mathrm{c}}{1+z} \frac{\sin(k x_{i})}{k},
\end{align}
\begin{align}
    \rho = \frac{\rho_{0}}{1-\frac{1+z_\mathrm{c}}{1+z} \cos(k x_\mathrm{i})},
\end{align}
\begin{align}
    v_\mathrm{pec} = - H_{0} \frac{1+z_\mathrm{c}}{\sqrt{1+z}\frac{\sin(k x_{i})}{k}} \mathbf{\hat{x}},
\end{align}
\begin{align}
    T = T_\mathrm{i} \left(\frac{1+z_\mathrm{c}}{1+z}\right)^{2} \left(\frac{\rho(x,z)}{\rho_{0}}\right)^{2/3},
\end{align}
with the unperturbed position x, the critical density of the Universe $\rho_{0}$, the Hubble constant $H_{0} = 100$ km s$^{-1}$ Mpc$^{-1}$, $z_\mathrm{c}=1$ and initial temperature T$_\mathrm{i} = 100$K. The test is realized at the very low resolution of $32^3$ particles. Additionally, we adopt an initial seed field in the y-direction of $10^{-3}$ physical G at redshift 100 ($9.80 \times 10^{-7}$ co-moving G). The test itself is unspectacular as the field is simply following the density perturbation. We show the results of the test in Fig.~\ref{fig:zeldovich}. We note that the cleaning scheme has almost zero effect in this particular test. We show the peculiar (x component) of the velocity field on the left, the magnetic field in the center panel, and the divergence error on the right for runs without cleaning (top row), with the fiducial cleaning (center row) and the fiducial cleaning where we adopt a cleaning speed that is a factor of two faster than the fastest wave speed. In the latter case, we see some minor reduction in the divergence error, but the effect is minor compared to the fiducial cleaning and no cleaning cases. However, the test is useful to test if the conversion of the cleaning scheme to co--moving units has been carried out correctly, which appears to be the case.


\bsp    
\label{lastpage}
\end{document}